\documentclass[twocolumn,apj]{emulateapj}
\usepackage{hyperref}
\usepackage{amsmath}
\usepackage{graphicx}
\usepackage{color}

\newcommand\editremark[1]{ {\color{red} #1}}

\newcommand\optional[1]{}

\newcommand\unit[1]{\, {\rm #1}}
\newcommand\mc{ {{\cal M}_c}}

\newcommand\Dh{ {D_{\rm H}}}
\newcommand\Dv{{D_{\rm v}}}      
\newcommand\Cv{{D_{\rm bns}}}

\newcommand\avq{{\left<q\right>}}
\newcommand\snrcut{\rho_c}    
\newcommand\snrcutPreferred{8}
\newcommand\snrsymbol{\rho}
\newcommand\mynmin{3000}
\newcommand\myratio{1\times 10^9}

\newcommand\tcutChirpMass{10^5 \unit{Myr}}
\newcommand\massTotalLimitInitial{40}

\newcommand\nsimsE{206}  
\newcommand\nsimsS{282}  

\newcommand\abbrvPSgrbs{PS-GRB}
\newcommand\abbrvPSmoreconstraints{PSC2}

\newcommand\abbrvCBC{CBC}
\newcommand\usheader{Uncertainties and smoothing{} }
\newcommand\onlineOnlyCaveat{available in the online version of the Journal}
\newcommand\orderof[1]{#1, multiplied by a factor of order unity}

\begin{document}
\title{Binary compact object coalescence rates:\\
 The role of elliptical galaxies} 
\author{R.\ O'Shaughnessy}
 \affil{
 Center for Gravitational Wave Physics, Penn State University,
 University Park, PA 16802, USA}
\email{oshaughn@gravity.psu.edu}
\author{ V.\ Kalogera}
\affil{Northwestern University, Department of Physics and Astronomy,
   2145 Sheridan Road, Evanston, IL 60208, USA}
\email{vicky@northwestern.edu}
\and
\author{Krzysztof Belczynski\altaffilmark{1,2,3}}
\affil{
     $^{1}$ Los Alamos National Laboratory, CCS-2/ISR-1 Group\\
     $^{2}$ Oppenheimer Fellow\\
     $^{3}$ Astronomical Observatory, University of Warsaw, Al.
            Ujazdowskie 4, 00-478 Warsaw, Poland
}
\email{            kbelczyn@nmsu.edu}


\begin{abstract}
In this paper we estimate  binary compact object merger detection rates for LIGO, including the potentially significant contribution from binaries that are produced in elliptical galaxies near the epoch of peak star formation.   Specifically, we convolve hundreds of model realizations of elliptical- and spiral-galaxy population
syntheses with a model for  elliptical- and spiral-galaxy star formation history as a function of redshift. Our results favor local merger rate densities of
$4\times 10^{-3} \unit{Mpc}^{-3}\unit{Myr}^{-1}$  for binary black holes (BH),
$ 3\times 10^{-2} \unit{Mpc}^{-3}\unit{Myr}^{-1}$  for binary neutron stars (NS), and $10^{-2}
\unit{Mpc}^{-3}\unit{Myr}^{-1}$ for BH-NS binaries.   We find that mergers in elliptical galaxies are a significant
fraction of  our total estimate for BH-BH and BH-NS detection rates; NS-NS detection rates are likely dominated by the
contribution from spiral galaxies. Limiting attention to elliptical- plus only those spiral-galaxy models  that
reproduce current observations of Galactic NS-NS, we find  slightly higher rates for NS-NS and largely similar ranges
for BH-NS and BH-BH binaries. Assuming a detection signal-to-noise ratio threshold of 8 for a single detector  (in
practice as part of
a network, to reduce its noise), corresponding to  radii $\Cv$ of the
effective volume inside of which a single LIGO detector could observe the inspiral of two $1.4 M_\odot$ neutron stars of
14\,Mpc and 197\,Mpc, for initial and advanced LIGO, we find event rates of {\em any} merger type of
$2.9\times 10^{-2}$ -- $ 0.46$ and $25-400$ per year  (at 90\% confidence level), respectively.
We also find  that the probability
$P_{\text{detect}}$ of detecting one or more mergers with this single detector can be approximated by 
(i) $P_{\text{detect}}\simeq 0.4+0.5\log
(T/0.01\unit{yr})$, assuming $\Cv=197 \unit{Mpc}$   and  it operates for $T$ years, for $T$ between $2$ days and
$0.1\unit{yr}$; or by (ii) 
$P_{\rm detect}\simeq 0.5 + 1.5 \log ( \Cv/32\unit{Mpc})$, for
one year of operation and for $\Cv$ between 20 and 70 Mpc. 
\end{abstract}

\maketitle

\section{Introduction}
Over the last decade the question of the Galactic inspiral rate of
binaries with two compact objects (neutron stars NS or black holes BH)
has attracted attention primarily because of the development and
planning of gravitational-wave interferometric detectors both on the
Earth and in space (e.g., LIGO and GEO600, described in \citet{gw-detectors-LIGO-original}; Virgo, described at the Virgo project
website \nocite{gw-detectors-VIRGO-website} \nocite{gw-detectors-VIRGO-original} {\tt www.virgo.infn.it}; and LISA, at
{\tt lisa.nasa.gov}).   These rate
estimates have been widely used in the assessment of gravitational
inspiral detectability, given assumed instrument
sensitivities.   A number of different groups have calculated inspiral
rates using population synthesis calculations, most commonly with
Monte Carlo methods 
\citep{1998ApJ...496..333F,1998AA...332..173P,
1994ApJ...423..659B,1999ApJ...526..152F,
StarTrack,ADM:Vos2003}.  Such studies consider the complete formation history
of double compact objects through long sequences of binary evolution
phases, terminated  by a gravitational-wave driven inspiral towards merger.
Though less critical for studies of the Milky Way, contributions to the present-day merger rate per unit volume from
early star formation has most often been ignored in merger rate 
calculations.   Since long inspiral delays before merger are not sufficiently uncommon, even for double neutron stars, merger
detection rate calculations should
account for the full time history of star formation, including star formation in the early universe; see for example
\citet{Regimbau2006-ellipticals}.  
 Additionally, our current understanding of single and binary star
evolution is incomplete.  
The many uncertaintites have been parameterized and the resulting
parameter space explored in some studies to determine 
the range of plausible results \citep[see,e.g.,][]{StarTrack2},
both by focusing narrowly on physically-motivated regions of parameter
space as in \citet{StarTrack2} and by broadly exploring all plausible population synthesis
simulations \citep[see,e.g.,][and references therein; henceforth
denoted \abbrvPSgrbs]{PSgrbs-popsyn}.
Finally, by assigning equal prior likelihood to any population
synthesis model, \abbrvPSgrbs{} estimated the relative likelihood of
any BH-NS or NS-NS merger rate.%
\footnote{Previously, \citet{PSconstraints,PSmoreconstraints} estimated LIGO binary merger detection rates
  by extrapolating from the local Milky Way based on the blue light density in the nearby Universe.  As we discuss here, their  estimate is
  accurate for short distances and for binary merger progenitors that are not
  preferentially long-lived (i.e., the median delay between birth and merger is less than $1 \unit{Gyr}$).} 

The population of binary black holes, however, behaves qualitatively
differently from BH-NS or NS-NS binaries: 
 (i) most of those that merge
do so only after a few to several $\unit{Gyr}$ delay after their birth
as binary stars; and  (ii) those BH-BH binaries with merger times from $1-10\unit{Gyr}$ have masses quite different from
the typical assumption of a $10 M_\odot$ BH; see Appendix.
The rarity,  long delay times, high masses and thus 
cosmologically significant  detection horizons
of binary black holes  have already been discussed in the literature 
(see,e.g., \citet{2006AA...459.1001K} and references therein).   However,  population
synthesis simulations extensive enough 
to contain a statistically meaningful binary black hole population come at
a significant computational cost.  
Even with adequate population synthesis simulations, given the long-lived nature of the progenitors, BH-BH
merger and detection rates depend critically on the earliest and least
certain star formation rates.  
Certain exotic but not exceptionally uncommon binary
evolution scenarios could lead to a ``high rate tail'' -- a scenario
with rare massive mergers, for example, that could plausibly produce a
merger detection in the near future.  In this paper we strive to
understand what realistic scenarios could lead to a ``high rate tail''
that LIGO could detect or, conversely, constrain.

Given the technical challenges and significant uncertainties involved,
only one paper \citep{PSconstraints} has previously estimated the relative likelihood of different
BH-BH  detection rate estimates expected from isolated binary star evolution, only  for the Milky Way.
In the present study we calculate the expected detection rate for
LIGO while for the first time simultaneously including (i) all past star
formation, particularly the overwhelming importance of elliptical
galaxies; (ii) a large database of results that accounts for the
dominant model uncertainties; and (iii) a careful treatment of the
mass distribution of BH-BH binaries.   
Additionally,  unlike previous analyses presenting distributions of
merger rates, we have post-facto varied the birth binary fraction of stars
from $100\%$ to $15\%$; this addition allows us to better
compare our results with the fiducial population synthesis model and
BH-BH merger rates presented in \cite{ChrisBH2007}.
Finally, whereas previous comparisons relied essentially on confidence intervals  (e.g., \abbrvPSgrbs{} and
\cite{PSmoreconstraints}, henceforth denoted \abbrvPSmoreconstraints), in this paper we generate true posterior distributions: each model is weighted by its relative
conditional likelihood given observations of merging NS-NS binaries in the Milky Way.

To fully assess the \emph{total} probability for a LIGO detection,
rather than limit attention to BH-BH mergers we explore the rate at
which all LIGO-detectable binaries (BH-BH, BH-NS, and NS-NS; 
henceforth collectively denoted as compact binary coalescences or \abbrvCBC{}s) merge
through gravitational wave emission.
As the methods used to determine LIGO detection rates from population
synthesis simulations are already extensively discussed in the literature
\citep[see,e.g.][and \abbrvPSgrbs]{2004MNRAS.352.1372B}, our presentation
only reviews those tools, emphasizing unique features of our present analysis (i.e.,
our methods to estimate BH-BH masses).
Specifically, in \S~\ref{sec:ligo}  we briefly
review the ingredients that enter into an estimate of the
gravitational wave detection rate for a short-ranged network ($z_{max}\ll 1$). While the local universe is
  emphasized in the text, 
networks of advanced ground-based interferometers can detect optimally
oriented BH-BH binaries with component masses $M\simeq 10-15
M_\odot$ at cosmologically significant distances $z\ge 0.1$.  
Merger rates on the past light cone of a detector become
inhomogeneous (versus redshift) at this scale, simply because the star formation rate increases dramatically
near $z\simeq 1-2$ during the epoch of galaxy assembly.  At these distances, the detection
rate must be  integrated over the mass distribution,
the full networked orientation-dependent sensitivity, and redshift-dependent merger rate.  This integral becomes increasingly sensitive to rare,
high-mass events that can be seen in the ancient universe. 
Therefore,  we generalize the short-range method described in \S~\ref{sec:ligo}
 to include all sources on the past light cone, to arbitrary redshift and with accurate
  orientation-dependent sensitivity.   In
\S~\ref{sec:sfr} we review the model for star formation in the universe adopted here.
 This experience is applied in \S~\ref{sec:popsyn} (for BH-BH
 binaries)
and \S~\ref{sec:popsyn:nonbh} (including NS-NS and BH-NS binaries too),
where we review our population synthesis calculations; and
 explain what features of those calculations influence our
predictions for the relative likelihood of a LIGO detection.  These
sections also describe Figure \ref{fig:ligo:rates}, which contains our
predictions regarding initial and advanced LIGO detection rates.
Finally, in \S~\ref{sec:constraints} we show how rate constraints from the observed sample of Milky 
Way NS-NS binaries affect the predictions for LIGO detection rates.

To summarize, by fully simulating the past history of the  local universe, this paper develops models for the present-day detection
rate of short-range ($z\ll 1$) and long-range ($z$ not much less than 1)
gravitational-wave detectors.
Our results are rate distributions, where each distribution includes  some normalization uncertainties (star formation
rate and fraction of stars born in binaries);
certain population synthesis model parameters; and our simulation Monte Carlo uncertainty. 
Further, our Bayesian approach to model constraints surveys many key  uncertainties that should be included, whatever the model family
involved,  when attempting to interpret upper limits or detections from pulsar populations and gravitational wave
observatories.
We believe our distributions represent the best predictions of a concrete, conservative model family
that includes both elliptical and spiral star formation yet is also  consistent with initial LIGO upper limits on
\abbrvCBC{} merger rates.

\section{Gravitational wave detection}
\label{sec:ligo}

This section primarily reviews  the final stage in any calculation of a LIGO
detection rate: the connection between, on the one hand, the event
rate per comoving volume ${\cal R}$ and mass distribution $p(m_1,m_2)$ of merging binaries
and, on the other hand, the detection rate $R_D$
[Eq. (\ref{eq:def:DetectionRate:Nearby})].   This
relation depends critically on the range to which LIGO could observe
each merger.
A thorough discussion of the LIGO range requires careful review of
data analysis strategies and interferometer network geometry, and
remains substantially beyond the scope of this paper.  For details 
see \citet{LIGO-Inspiral-s3s4,LIGO-Inspiral-S5-1styear}  and references
therein.   For example, with modern multi-mass-region searches using
inspiral templates,  one
essential effect is a merger detection threshold in
SNR that changes with total binary mass, as the well-understood portions of 
detection templates grow shorter. 
In this paper, we adopt the customary set of approximations (described
below) and ignore
the errors they introduce.


The distance to which LIGO is sensitive can be calculated using the
sensitivity  of the detector (strictly, the strain noise spectral distribution $S_h$), the masses of the inspiralling bodies,
and the emitted waveform.   [For simplicity, we will assume the
black holes initially nonspinning.]  Specifically, for a source with
masses $m_1$ and $m_2$, merging at a luminosity distance $D$ 
(or equivalently at redshift $z$, or comoving distance $r$)
with a binary inclination $\iota$ and a relative orientation to the
detector given by $\theta,\phi$ (the orientation on the sky) and
$\psi$ (a phase angle in the plane of the detector), the signal-to-noise seen by
a matched-filter search in the LIGO data stream can be expressed 
relative to the strain noise spectrum of a single interferometer $S_h$
and the Fourier transform $\tilde{h}(f)$ of gravitational wave strain $h(t)$ received at
the detector by an  arbitrarily oriented and located source by
\begin{eqnarray}
\label{eq:def:LIGORangeSNR}
\snrsymbol^2&=&4\frac{w(\theta,\phi,\psi,\iota )^2}{D^2} 
\int_0^{\infty} df \,
 \frac{|r \tilde{h}(m_1(1+z),m_2(1+z),f)|^2}{S_h(f)} \nonumber \\
\end{eqnarray}
using units with $G=c=1$, the known dependence of planar-symmetry
gravitational waves on angle, and a standard one-sided Fourier
convention for $\tilde{h}$ and $S_h$ versus frequency $f$.
The function  $w(\theta,\phi,\psi,\iota)$ takes values between 0 and 1 and
completely encompasses 
the detector- and source-orientation dependent
sensitivity for the most common sources, those that are  dominated by
$l=m=2$ quadrupole emission\footnote{Peak sensitivity is attained for a source
  directly overhead with orbital plane normal to the line of sight.}
%
\begin{eqnarray}
\label{eq:def:w}
w(\theta,\phi,\psi,\iota)&=&
  \sqrt{F_+^2 (1 + \cos^{2} \iota)^2/4+ F_{\times}^2\cos^{2} \iota} \\
F_\mathrm{+} &=&
 \frac{1}{2}(1+\cos^{2} \theta) \cos 2 \phi \cos 2 \psi
\nonumber\\ &+& \cos \theta \sin 2 \phi \sin 2 \psi \, \\ 
F_\mathrm{\times} &=&- \frac{1}{2}(1+\cos^{2} \theta) \sin 2 \phi \cos 2 \psi
\nonumber \\ &+& \cos \theta \cos 2 \phi \sin 2 \psi
\end{eqnarray}
To a rough approximation that depends on the data analysis strategy
used and the amount of nongaussian noise present in the detector, a
single LIGO interferometer can detect 
the gravitational wave signature of a merging binary if $\snrsymbol >
\snrcut$  (see, e.g., \citet{LIGO-Inspiral-s2-bbh,LIGO-Inspiral-s3s4},
 and references therein; henceforth we adopt $\snrcut=\snrcutPreferred$).  For this
reason, the previous expression for the signal to noise
[Eq. (\ref{eq:def:LIGORangeSNR})] is often re-expressed as
\begin{eqnarray}
\label{eq:def:Dh}
\snrsymbol &=& 
\snrcut  w(\theta,\phi,\psi,\iota) \frac{\Dh(m_1(1+z),m_2(1+z))}{D} 
\end{eqnarray}
which, by comparison with Eq. (\ref{eq:def:LIGORangeSNR}), implicitly
defines  the  \emph{horizon distance} $\Dh$ -- the maximum
luminosity distance to which the detectors are
sensitive -- as a function of the redshifted masses $m_1(1+z)$
and $m_2(1+z)$ of the binary
(see, e.g.,   \citet{Anderson:2001}, \citet{1993PhRvD..47.2198F} and \citet{1998PhRvD..57.4535F} for a brief
review of the theory underlying the direction-dependent LIGO
sensitivity as well as for expressions that approximate $\Dh(m_1,m_2)$
for low-mass \abbrvCBC{}  binaries\footnote{A more thorough review of the orbital geometry and
  waveforms underlying the reach of a single-interferometer search for
  circularly inspiralling point particles can be found in 
\citet{DuncanBrownThesis} and (for the more complex case of spinning
binaries) in 
\citet{ACST,BuonannoChenVallisneri:2003b}.  Additionally, 
\citet{LIGO-Inspiral-s3s4-Galaxies} discusses how
this orientation-dependent sensitivity influences the present-day LIGO
search.
}).

  More commonly used is the \emph{volume-averaged distance}
$\Dv$, chosen so the volume of a sphere of radius $\Dv$ agrees with
the average volume enclosed in  a (Euclidean) detection surface:
\begin{eqnarray}
 \Dv^3&=& \frac{1}{4\pi}\Dh^3
\int d\Omega \frac{d\psi}{\pi} \frac{d \cos \iota}{2} w^3
 \nonumber \\
\label{eq:def:Dv}
 &\simeq&  (\Dh/2.26)^3
\end{eqnarray}
The volume averaged distance is a meaningful measure of sensitivity
only when the horizon range is much smaller than the Hubble scale.

A  \emph{network} of detectors can  coherently  add signals from each
interferometer to increase  the  signal-to-noise $\rho$ associated with each signal and by implication its
  reach.
To a first approximation, ignoring small differences
in sensitivity due to both their intrinsic differences and their  orientations,\footnote{%
The small bias ($\simeq 10\%$) introduced by this approximation is
much smaller than the characteristic
uncertainties discussed later in this paper;  see, for example, 
 LIGO site
  location information in
LIGO-T980044-10 (available at
\href{http://admdbsrv.ligo.caltech.edu/dcc/}{http://admdbsrv.ligo.caltech.edu/dcc/}) and the LAL software
documentation (available from
\href{http://www.lsc-group.phys.uwm.edu/daswg/projects/lal.html}{http://www.lsc-group.phys.uwm.edu/daswg/projects/lal.html}).
%
}
 the combined SNR  all
interferometers is higher by roughly  $\sqrt{\sum_k L_k^2/L_1^2}$ where $L_k$ is the length of
the $k$th interferometer.  For example, this factor is approximately
$g_N\simeq \sqrt{1+1+1/4}\simeq \sqrt{2.25}$ for the initial LIGO network, consisting of two $4\unit{km}$ and one $2\unit{km}$
interferometer;  see 
\cite{CutlerFlanagan:1994} for details on realistic multidetector beampatterns..
However, in part because a network performs more trials of the same data streams (e.g., for each sky position) and is
sensitive to both polarizations simultaneously, for the same false alarm rate the network detection threshold
$\rho_{c,net}$ is generally greater than the threshold $\rho_c$   for the individual detectors; see for example
\citet{CutlerFlanagan:1994} and \citet{2003PhRvD..67j4025B}
for a discussion in the case of gaussian noise.
One can therefore speak of a single-interferometer ($\Dh$) and network ($\Dh^{(n)}\simeq
g_N\Dh (\snrcut/\rho_{c,net})$)  horizon distance.   Unfortunately, only expert analysis of real data
can  determine the sensitivity of real gravitational-wave networks, not the least because  noise
statistics (and therefore $\rho_{c,net}$) are highly  detector- and even search-dependent.
To avoid ambiguity and  misleading approximations, we  provide results for {\emph
  single},  idealized interferometers  assuming a detection threshold of $\snrcut=8$  (e.g., appropriate to    gaussian noise).  This
  reference sensitivity agrees with the customary sensitivity measure used in the LIGO coalescence search, a threshold
  of $\snrcut=8$ for a single detector as part of a network.\footnote{For the real LIGO interferometers, no single
    detector search at this threshold is possible;  multi-detector coincidence is required to reduce the nongaussian
    background to roughly the level indicated (hence ``as part of a network'').
    }
For sources in the local universe, without loss of
generality the reader can scale our results to any realistic network and search.

For sources of sufficiently
low mass ($M_{tot}\lesssim 20 M_\odot$), LIGO is primarily sensitive to less-relativistic and
therefore better-understood phases of the spiral-in, allowing us to
approximate the waveform by its 
Newtonian limit \citep{Peters:1964}.  After considerable algebra
\citep[see,e.g.][]{1998PhRvD..57.4535F,CutlerFlanagan:1994}, the  signal-to-noise
due to an optimally-oriented inspiral at comoving distance
$r$ (corresponding to a redshift $z(r)$) can be related to the strain 
sensitivy $S_h$ by   
\begin{eqnarray}
\label{eq:def:LIGORangeSNR:Lowmass}
\snrsymbol^2
 &=&  \frac{5 w^2 [\mc(1+z)]^{5/3}}{6 \pi^{4/3} D^2} \int_0^{\infty}  df \,
\frac{{ f}^{-7/3}}{S_h(f)}
\end{eqnarray}
where $\mc\equiv (m_1 m_2)^{3/5}/(m_1+m_2)^{1/5}$ is the ``chirp
mass'' of the binary.
 Based on this expression,  individual interferometers in 
the 
initial\footnote{We use a published LIGO S5 sensitivity  (for the  LHO interferometer, on March 13th, 2006), available
  at
  \href{http://www.ligo.caltech.edu/\~jzweizig/distribution/LSC_Data/}{http://www.ligo.caltech.edu/\~{}jzweizig/distribution/LSC\_Data}.
}
, and advanced\footnote{ We adopt an advanced LIGO noise curve from LIGO T0900288, available as
{\tt ZERO\_DET\_high\_P.txt} at
\href{https://dcc.ligo.org/DocDB/0002/T0900288/002/}{dcc.ligo.org/DocDB/0002/T0900288/002/}.
This noise curve is taken from the GWINC program with Mode 1b parameters, as described in
LIGO document LIGO-T070247-01-I: 125 W input laser power, 20\% signal recycling mirror (SRM) transmissivity,
and no detuning of the signal recycling cavity.
}
LIGO networks can see \emph{low-mass} binaries ($m_1+m_2 \lesssim \massTotalLimitInitial M_\odot$)\footnote{For
    comparison, full strong-field numerical relativity waveforms suggest a single-interferometer initial LIGO horizon distance $\Dh$ to the most massive equal-mass binaries
    $M\simeq \massTotalLimitInitial$
    that differs from the low-mass limit by \orderof{$2\%$}; see \cite{gwastro-Ajith-AlignedSpinWaveforms}.
   On the contrary, our limitation to \emph{equal-mass, nonspinning} binaries introduces greater error: for binaries
   $M\le 16 M_\odot$, mass ratio
   corrections are roughly $4\%$, while black hole spin-dependent corrections are  \orderof{$0.1 J/m_{bh}^2$} (if
   aligned).  At the extreme, BH-NS binaries with strong spin-orbit misalignment and thus precession could be visible only out to
   \orderof{$0.5\Dh_{ref}$}, for $\Dh_{ref}$ the horizon distance to a comparable aligned binary.  Such extreme uncertainties
   in range  are less plausible, as isolated binary evolution models
   favor spin-orbit alignment.
}  out to a horizon (volume-averaged) distance
\begin{equation}
\label{eq:def:LIGORange:ShortRangeSensitivity}
D_{\rm H(v)} \simeq {\cal C}_{\rm H(v)} (\mc/1.2 M_\odot)^{5/6}
\end{equation}
where
${\cal C}_{\rm H(v)}=31 (14) \unit{Mpc}$  and ${\cal C}_{\rm H(v)}=445 (197) \unit{Mpc}$ respectively.   For clarity, the distance ${\cal C}_v$ will henceforth be denoted $\Cv$, the
volume-averaged distance to which a binary neutron star inspiral can be detected by a single interferometer as part of a network.
In terms of this  horizon distance $\Dh$, the intrinsic present-day rate of events
per unit comoving volume ${\cal R}(0)$, and the chirp mass distribution of
merging binaries $p(\mc)$, the average rate of events with 
$\snrsymbol > \snrcut$ occurring in the nearby universe can be expressed as 
\begin{eqnarray}
\label{eq:def:DetectionRate:Nearby}
R_D 
 = {\cal R}(0)&&\int_{\frac{w\Dh}{D}>1} r^2 dr d\Omega
          \frac{d\psi}{\pi}
          \frac{d\cos \iota }{2} 
 p(\mc)       d \mc
\\
\label{eq:def:DetectionRate:Local}
= {\cal R}(0)   &&
 \left[\int d\mc \frac{4\pi}{3}\Dv(\mc)^3
  p(\mc)  \right]\, .
\end{eqnarray}
where 
the integral is taken over all detectable
combinations of masses, orientation, and location. 
For  low-mass binary mergers  ($M< \massTotalLimitInitial M_\odot$), for which Eq. (\ref{eq:def:LIGORange:ShortRangeSensitivity}) applies, the chirp mass-weighted average
simplifies to
\begin{eqnarray}
\label{eq:def:DetectionRate:Local:LowMass}
R_D&\simeq& {\cal R}(0)  \frac{4\pi ( {\cal C}_v)^3}{3} 
 \int p(\mc)  (\mc/1.2 M_\odot)^{15/6} d \mc\; ,
\end{eqnarray}
an expression that has been applied extensively in the literature 
 to translate event rates per unit volume into LIGO detection rates
 \citep[see,e.g.][and references therein]{PSmoreconstraints}.  

\subsection{Rates for advanced detectors}
\label{sec:aLIGO:RateMethods}

As the product of the present-day merger rate and an effective detection volume, the detection rate estimate
Eq. (\ref{eq:def:DetectionRate:Nearby}) assumes a locally homogeneous and isotropic universe.
Advanced detectors can see back to sky-position-dependent epochs with noticeably different star formation history
($z>0.1$).  
%
%
For advanced detectors, a more generic expression must be used, which   integrates over the mass distribution,
the redshift-dependent merger rate, and the  orientation-dependent reach $w$.

For the purposes of this paper, we continue to assume the maximum luminosity distance to which a single interferometer
can identify a merger in noise  is well-described by the inspiral phase  [Eq. \ref{eq:def:LIGORangeSNR:Lowmass}].  To normalize
our estimate, we adopt a single-interferometer advanced LIGO range of $445\unit{Mpc}$.    Based on
the optimal range for a given chirp mass, each simulation's the chirp mass distributions, and the single-interferometer beampattern $w$ [Eq. \ref{eq:def:w}],  we
determine the fraction of all mergers at redshift $z$ that could be detected, $P_{ok}(z)$:
\begin{eqnarray}
P_{ok}(z)&=& \int_{D(z)<w \Dh}
          \frac{d\Omega}{4\pi}
          \frac{d\psi}{\pi}
          \frac{d\cos \iota }{2}  
       d\mc p(\mc)
\end{eqnarray}
The overall single-interferometer detection rate is therefore the sum over the past light cone of the (redshifted) rate of mergers on it,
times the fraction $P_{ok}$ of mergers that could be detected:
\begin{eqnarray}
R_D &=& \int dz \frac{dV}{dz} \frac{{\cal R}(t)}{1+z} P_{ok}(z) 
\end{eqnarray}
For a maximum reach $z\ll 1$, these two expressions are equivalent to Eq. \ref{eq:def:LIGORangeSNR:Lowmass}.

\optional{
\subsection{Discussion: LIGO errors}
\editremark{TENTATIVE}

- mass range validity

- spin error

- mass ratio error

..can mostly be tabulated for these systems.
}

\section{Multicomponent Star formation History}
\label{sec:sfr}
The LIGO detection rate depends  on  the mass distribution of 
merging binaries $p(\mc)$ and on the
average rate of
mergers per unit comoving volume, ${\cal R}(t)$.  This rate, in turn,
generally depends on the net contribution from 
multiple star-forming populations, a contribution that convolves the
star formation rate in each component (the subject of the present
section) with the rate at which each components' star-forming gas
yields \abbrvCBC{} mergers (the subject of \S~\ref{sec:popsyn}).
More specifically, the merger rate density 
 ${\cal R}(t)$, a quantity required to calculate the detection
rate, is obtained from (i) the star formation rate  in each component   $d\rho_{C}/dt$
where $C$ indexes the different star-forming types, $C=e$ for ellipticals and $s$ for spirals, 
defined as the mass per unit comoving volume and time that forms as stars; (ii) the
mass efficiency $\lambda_{C}$ at
which each type of \abbrvCBC{} binaries form, defined as the total number of binaries
that survive isolated evolution and form \abbrvCBC{} binaries per unit star
forming mass; and (iii) the
probability distribution $dP_{m C}/dt$ for merger events to occur
after a delay time $t$ after star formation,  where $dP_{m C }$ is the fraction of mergers occuring between $t$ and
$t+dt$ after the progenitor binary forms at $t=0$:
\begin{eqnarray}
\label{eq:comovingRateNet}
{\cal R}(t) &=& \sum_{C}{\cal R}_{C}(t) \\
\label{eq:int:comovingRate}
{\cal R}_{C}(t) &=& \int_{-T}^t dt_b \lambda_{ C}\frac{dP_{m C}}{dt}(t-t_b) \frac{d\rho_{ C}}{dt}(t_b)
\end{eqnarray}
 where the integration variable $t_b$ is time at which $\lambda  d\rho$ binaries per unit volume are born.
As first recognized by
\citet{Regimbau2006-ellipticals} and as demonstrated systematically in
\abbrvPSgrbs, 
 a multicomponent star formation
history that includes both elliptical and spiral galaxies must be
applied when calculating present-day merger rates of double compact
objects.  Simply put, even though elliptical galaxies formed stars long
ago, $dP/dt$ decays slowly enough (roughly as $1/t$; see, e.g.,
\abbrvPSgrbs) that the high
rate of star formation in the early universe can lead to a
significant elliptical-galaxy-hosted merger rate density in the
present-day universe.  Because elliptical galaxies have distinctly
different star forming conditions (e.g., metallicities) than
present-day star-forming galaxies, models for stellar evolution in the
universe should not require identical behavior in present-day
solar-metallicity galaxies and in young ellipticals.
A full multicomponent treatment is necessary.  The first 
multicomponent star formation history was applied to predict  BH-NS and NS-NS
merger rates in \abbrvPSgrbs; this paper employs the same
framework and preferred star formation history, discussed below.

\begin{figure}
\includegraphics[width=\columnwidth]{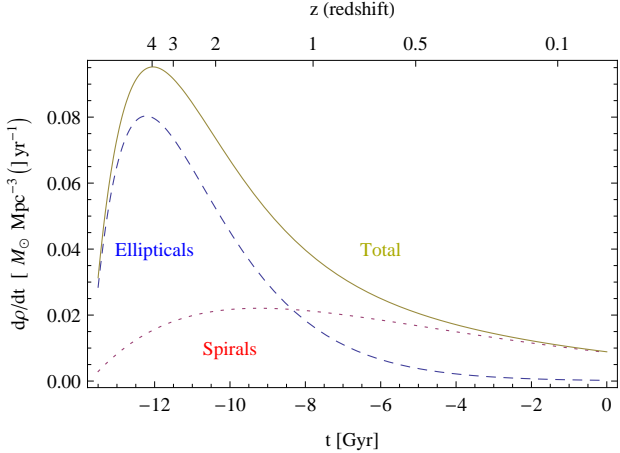}
\caption{\label{fig:sfr:observed} 
 Star formation history of the universe used in this paper versus time, relative to the present day.
Solid line:  Net star formation history implied by
  Eq. (\ref{eq:sfr:model:twocomponent}).
Dashed, dotted line: The star formation history due to  elliptical and
spiral components.
}
\end{figure}

We adopt the two-component star formation history
model presented by \citet{NagamineTwoComponentSFR2006}.  This model
consists of an early ``elliptical'' component and a fairly steady
``spiral'' component, with star formation rates given by 
\begin{eqnarray}
\label{eq:sfr:model:twocomponent}
\dot{\rho} &=&\dot{\rho}_e + \dot{\rho}_s \\
\dot{\rho}_{C}&=& A_{C} (t/\tau_{C}) e^{-t/\tau_{C}}
\end{eqnarray}
where cosmological time $t$ is measured starting from the beginning of
the universe
and where the two components  decay 
timescales are $\tau_{e,s}=$ 1.5 and 4.5 Gyr, respectively \citep[see Section 2
and Table 2 of ][]{NagamineTwoComponentSFR2006}.  
These normalization constants $A_{e,s}= 0.22, 0.06 M_\odot{\rm
  yr}^{-1}{\rm Mpc}^{-3}$ were chosen by \citet{NagamineTwoComponentSFR2006} so the
integrated amount of elliptical and spiral star formation reproduce
the present-day census of baryonic matter in ellipticals and spirals,
allowing for a certain fraction of gas recycling between different
generations of stars.\footnote{Our previous paper \abbrvPSgrbs{} incorrectly listed a larger value  $A_s=0.15
  M_\odot{\rm yr}^{-1} {\rm Mpc}^{-3}$.  
}
Figure
\ref{fig:sfr:observed} illustrates the star formation rates assumed by
the model adopted here.

Each component forms stars in its own distinctive conditions, set by
comparison with observations of the Milky Way and elliptical galaxies.
We assume mass converted into stars in the fairly steady ``spiral''
component with
solar metallicity and with a fixed high-mass initial mass function (IMF) power law [$p=-2.7$
in the broken-power-law Kroupa IMF; see
\cite{KroupaClusterAverageIMF2.7}].  
On the other hand, we assume stars born in the   ``elliptical'' component are
drawn from a
broken power law IMF with high-mass index within  $p\in[-2.27,
-2.06]$ and metallicity $Z$ within $0.56<Z/Z_\odot<1.5$.
These elliptical
birth conditions agree with observations of both old ellipticals in the
local universe \citep[see][and references therein]{astro-ph..0605610} as well as of young starburst clusters
\citep[see][and references therein]{2005ApJ...631L.133F,ADM:Zha99}.

Since  two independent
stellar populations  give birth to \abbrvCBC{}s,
the net detection rate $R_D$ is the sum of the detection rate
due to ellipticals and spirals.   In particular,  the probability
distribution function 
for $R_D$ is necessarily the convolution of the distribution functions
for detection rates due to elliptical and spiral galaxies,
respectively.     Because of the many orders of magnitude uncertainty
in merger rates and the need to smooth across relative errors, however, all our PDFs are stored and
smoothed in a logarithmic representation, as $p(\log R_D)$; see for
example \abbrvPSgrbs.  For this reason, the convolution takes on an
unusual form, involving an integral over the fraction $f$ of the
detection rate due to ellipticals (i.e., $f = R_{D,e}/R_D$).  Using
this variable to form the logarithmic convolution, 
the distribution for the net rate $\log R_D$ can 
be calculated from the distributions for ellipticals and spirals
($p_{e,s}$) as 
\begin{eqnarray}
\label{eq:logconvolve}
p(\log R_D) &=& \int_0^1\frac{df}{f(1-f)\ln(10)} 
\nonumber\\ &\times &
 p_e(\log R_D + \log f) 
\nonumber\\ &\times &
p_s(\log R_D + \log (1-f))\\
&\equiv & p_s \oplus p_e
\end{eqnarray}

\noindent \emph{Limitations of two-component model}:  Our two component model implicitly adopts two extremely strong
assumptions: (i) that metallicity in each component does not significantly evolve with redshift and (ii) that all star-forming gas in
each component is identical.   While sufficient for most problems involving binary evolution, these approximations
may be particularly ill-suited to determining the progenitors of gravitational wave detections.  Very low metallicity environments could produce
BHs with exceptionally high mass; also, the BH binary merger rate 
increases strongly with metallicity \citep[see,e.g.][]{popsyn-MaxMassBH-Chris2009}.     Because of the long typical delay between
progenitor birth
and BH-BH merger, very low-metallicity environments in the early universe could contribute significantly to the
present-day gravitational-wave detection rate \cite{popsyn-LowMetallicityImpact-Chris2008}.  
Even at present, low-metallicity environments exist often enough to potentially dominate the local merger rate
\citep[see,e.g.,][]{popsyn-LIGO-SFR-2008}.  For example, the nearby BH-BH progenitor binary IC 10 X-1 both lies in a low
metallicity environment and suggests a  high BH-BH detection rate for initial LIGO (\orderof{$0.5\unit{yr}$}, strongly
dependent on survey selection effects; see \cite{bhrates-Chris-IC10-2008}).

Despite strong theoretical and observational evidence for a significant low-metallicity contribution, we have neither
the confidence in the necessary 
astrophysical inputs (i.e., a redshift-dependent metallicity distribution) nor the required databse of simulations to
include these metallicity effects at this time.  Further, unless initial
LIGO detects BH-BH mergers, the peak detection rates cannot be significantly larger than what our model family
predicts; see Figure \ref{fig:ligo:rates}.    For this reason we continue to adopt the simple and conservative two-component mechanism
outlined above.

\noindent \emph{Relation to blue light density}: Previously, merger rate estimates for the Milky Way have been
scaled up to the local universe, assuming mergers track blue light  \citep[see,e.g.][and references
therein]{LIGO-Inspiral-s3s4-Galaxies}.   As roughly speaking blue light traces star formation 
\citep[see,e.g.][]{1984ApJ...284..544G,ADM:Ken98,popsyn-LIGO-SFR-2008}, this extrapolation roughly agrees with our
estimate for binary merger rates in spiral-like galaxies, with a suitable choice of normalization (i.e., star formation rate
per unit blue light); see the conclusions for a detailed numerical comparison.  \cite{popsyn-LIGO-SFR-2008} explains why blue light
normalization was not adopted in  this paper.

\section{\abbrvCBC{} Population synthesis models}
\label{sec:popsyn}
As outlined in \S~\ref{sec:ligo} and \S~\ref{sec:sfr} [see particularly Eqs.
(\ref{eq:def:DetectionRate:Nearby},\ref{eq:int:comovingRate})], in
order to estimate how often \abbrvCBC{} binary mergers could be seen
with gravitational-wave detectors we must know several features of the
\abbrvCBC{} population: 
(i) how many \abbrvCBC{} binaries
form from any given star forming mass (characterized by the mass
efficiency $\lambda$); 
(ii) how long these \abbrvCBC{} binaries last between birth as stars and
merger through gravitational wave emission (characterized by $dP/dt$);
and (iii) what masses these merging \abbrvCBC{} binaries have (characterized
by a probability distribution in chirp mass $p(\mc)$).   
 In the
absence of observations, these properties must be obtained from
theoretical models of stellar and binary evolution.
We study the formation of compact objects with the \emph{StarTrack}
population synthesis code, first developed by  
\citet{StarTrack} and recently significantly extended as described in
detail in \citet{StarTrack2}.    Since our understanding of the
evolution of single and binary stars is incomplete, this code parameterizes
several critical physical processes with a great many parameters
($\sim 30$), many of which influence compact-object formation
dramatically.  
In this specific study, in addition to the IMF and
metallicity (which vary depending on whether a binary is born in an
elliptical or spiral galaxy), seven parameters strongly
influence compact object merger rates: the supernova kick distribution
(modeled as the superposition of two
independent Maxwellians, using three parameters: one parameter for the
probability of drawing 
from each Maxwellian, and one to characterize the dispersion of each
Maxwellian), the stellar wind strength, the 
common-envelope energy transfer efficiency, the fraction of angular
momentum lost to infinity in phases of non-conservative mass transfer,
and the 
relative distribution of masses in the binary.     
For this reason, following methods largely outlined in our previous
work in \abbrvPSgrbs{} and  \abbrvPSmoreconstraints, 
we perform a broad \emph{parameter study} to
ensure  all plausible population synthesis models have been explored.
%
We include simulations of two
different classes of star-forming conditions: ``spiral'' conditions,
with $Z=Z_\odot$ and a high-mass IMF slope of $p=-2.7$, and
``elliptical'' conditions, with a much flatter IMF slope $p\simeq -2.3$ and a range
of allowed metallicities $0.56<Z/Z_\odot<1.5$.  
Though we have performed thousands of simulations of these two types
of conditions in the past (e.g.,\abbrvPSgrbs), to obtain better statistics on rare BH-BH
events,
we have
constructed a new set of archives of 
\textbf{$N_S\equiv \nsimsS$} and
\textbf{$N_E\equiv \nsimsE$} simulations of these two type of conditions, respectively
Within each set, we select
parameters randomly according to the distributions
described above; see the Appendix for more details. 
%

The model database described above is statistically very similar to the database  \abbrvPSgrbs{} previously used 
to estimate NS-NS and BH-NS merger rates, with one
critical exception: the simulations used here are extensive enough to have a statistically significant number of merging BH-BH
binaries. 
Merging BH-BH binaries are a small fraction of the
total BH-BH population and an exceedingly rare consequence of isolated
binary evolution.  For example, the number ($n$) of BH-BH binaries  in each of our simulations was always by
construction $3000$ and could be as high as few$\times 10^5$ (see Figure \ref{fig:popsyn:selection}).  On the contrary,
the number ($m$) of \emph{potentially merging} BH-BH binaries whose delay between formation
and merger is less than $13.5\unit{Gyr}$  is typically $\orderof{30}$; despite our efforts, a spiral-galaxy simulation produced only one  merging BH-BH binary   (Figure \ref{fig:popsyn:selection}). 
%
Further, among those few simulated binaries the most massive few can be seen to the greatest distance; thus, only a fraction
of these  $m$  binaries
significantly impact the average chirp mass $\left<\mc^{15/6}\right>$ [Eq. (\ref{eq:def:DetectionRate:Nearby})]
calculated therefore our estimate of the detection rate.
Given the small number statistics upon which we build our BH-BH detection rate predictions,  we very carefully quantify
and propagate uncertainties into our estimated BH-BH (and also  BH-NS, NS-NS) detection rates. 


\subsection{Model uncertainty and recent literature}
\label{sec:sub:ModelUncertaintyReview}


Our simulations do not explore some known model uncertainties that could significantly increase the predicted merger rate.   As
noted previously, our star forming conditions involve only near-solar metallicity
\cite{bhrates-Chris-IC10-2008,popsyn-LowMetallicityImpact-Chris2008,popsyn-MaxMassBH-Chris2009}.
Additionally, we assumed that the common-envelope evolution of a Hertzsprung gap donor led to stellar merger, not a
compact binary.  However, as discussed in \cite{ChrisBH2007}, the alternative and still plausible option will increase the BH-BH
merger rate by $\times 500$.  Finally, we assume binaries form through isolated evolution alone, not allowing for
any enhancement in young globular proto-clusters; see, e.g., \cite{clusters-2005} and \cite{2008ApJ...676.1162S}.
As mentioned previously, because these factors should \emph{increase} the detection rate, our results are the predictions of a conservative model family.

Our simulations also do not self-consistently explore all model uncertainties corresponding that are comparable or smaller than
the statistical simulation errors described extensively below.  For example, 
as the StarTrack code has evolved over the extended assembly of our archive and this paper, our simulations differ
somewhat from other contemporary papers that employ it.  Notably, unlike \cite{ChrisBH2007}, we adopt the full Bondi
accretion rate during common-envelope evolution; by trapping more mass in the binary, this generally leads to 
larger detection rates.\footnote{\label{foot:AlternateBondi} Though some black hole masses are influenced noticably by a change in mass accretion, 
this change does not always lead to dramatic changes in BH-BH merger- and \emph{detection} rates.  For example,  \cite{ChrisBH2007}
    provide a side-by-side comparison of the BH-BH merger rate for two different choices of 
    the accretion (models A,B in their tables 2,3).  They see less than a factor 2 change between 
    these two extreme limits, much less than the typical variation due to the parameters we explore. 
    [See also their Table 1 on BH-BH formation channels in these two models.]
    Given our limited computational resources, the fact that BH-BH detection rates are ubiquitously dominated
    by the highest-mass BH-BH binaries that merge (not the marginal BH-BH binaries where one formed through AIC), 
    and recongizing that other model parameters (NS birth mass distribution)
    and physical inputs (metallicity distributions; systematic SFR normalization and model error)
    will perturb our predictions by a comparable amount to the factor of 2 mentioned above, we are comfortable with neglecting 
    this parameter in this survey of BH-BH detection rates.
}
Also, the StarTrack code does not yet fully explore all uncertainties in single star evolution, such as
uncertainties in stellar radii \citep{1999ApJ...526..152F} and due to  rotational effects
(see, e.g., \cite{2009AA...497..243D} \cite{2001AA...373..555M} and references therein).
 Nor did we explore modifications to the recipes used by \emph{StarTrack} for single and binary evolution.   For
  example, StarTrack models orbital decay in   common-envelope  evolution
with  a single  factor $\alpha \lambda$, though in principle $\lambda$ differs between and
  can be calculated for 
  different donor star configurations.   For a supernova, the initial (pre-fallback) compact object remnant mass is
  estimated by tabulated estimates of the core mass; see \cite{StarTrack2}.  Compact remnants (post-fallback) more
  massive than $2.5 M_\odot$ are assumed to form black holes.

 Finally, all of our results are sensitive to the total number of massive progenitor stars and  our estimate for the high-mass IMF.
While our computationally limited exploration cannot give an idealized, fully-marginalized theoretical prediction, our approach is
extremely useful for the reverse problem:  how this particular
  concrete, conservative model family can be tested against future gravitational wave observations.


\subsection{Properties of all \abbrvCBC{} binaries}

Though the calculation was applied only to BH-NS and NS-NS mergers,
\abbrvPSgrbs{} described how to calculate two of the three essential ingredients needed to calculate the \abbrvCBC{}
detection rate [Eqs. (\ref{eq:def:DetectionRate:Nearby},\ref{eq:int:comovingRate})]:
(i) the number of \abbrvCBC{}  merger events per unit mass in progenitors (the
mass efficiency $\lambda_{C,\alpha}$);  and 
(ii) the probability  that given a \abbrvCBC{}
progenitor, a merger  occurs between $t$ and $t+dt$ since the binary's birth as
two stars (the delay time distribution  $dP_{c,\alpha}/dt$ ).   For example, as in \abbrvPSgrbs{} we estimate the mass efficiency $\lambda$ for forming a
merging binary of type $K$(=BH-BH,BH-NS,NS-NS)
from  the number  $n$ of binary progenitors of $K$ with
\begin{eqnarray}
\label{eq:popsyn:lambda}
\lambda &=& 
\frac{n}{N} \frac{f_{cut}}{\left< M \right>} 
\end{eqnarray}
where 
$N$ is the total number of 
binaries simulated, from which the $n$ progenitors of $K$ were drawn;
$\left<M\right>$ is the average mass of all possible binary progenitors; and 
$f_{cut}$ is a correction factor accounting for the great many very
low mass binaries (i.e., with primary mass $m_1<m_c=4 M_\odot$) not
included in our simulations at all.  Expressions for both
$\left<M\right>$ and $f_{cut}$ in terms of population synthesis model
parameters  are provided in Eqs. (1-2) of \citet{PSutil2}.  
We also estimate the delay time distribution as before, by smoothing (in $\log t$) the $n$ simulated delays $t$ between binary formation
and merger;  see particularly the Appendix of \abbrvPSgrbs{}.
Finally, though \abbrvPSgrbs{} does not mention  masses, we use precisely the same logarithmic smoothing technique
to estimate $dP/d\mc$ from a set of binaries; see our Appendix.

The procedures described above lead to very reliable results for NS-NS and BH-NS binaries, because many binaries $n$ and
even merging binaries $m$ are present in each simulation.
However, most simulations have only a few
BH-BH binaries whose delays between birth and merger are less than the age
of the universe.
With so few merging binaries per simulation,
statistical uncertainties would severely limit our ability to
determine the physically relevant portion of delay time  ($dP/dt$) and
chirp mass ($p(\mc)d\mc$) distributions based on those binaries
alone. 
However, since   gravitational wave decay depends sensitively
on the post-supernova orbital parameters of a newly-born BH-BH binary,
the population of \emph{merging} BH-BH binaries should be very similar to a
population with marginally wider orbits but often dramatically longer
decay timescales.
For this reason, in this paper we will improve our statistics for
$dP/dt$ and $p(\mc)$ by including nearly-merging binaries
with delay times $t$ between birth and merger less than a
 cutoff  $T=\tcutChirpMass$, as justified by our
 studies in the Appendix. 

\subsection{Varying the Binary Fraction}
\label{sec:smoothing:BinaryFraction}
In our population synthesis simulations we systematically varied
almost all parameters that could significantly impact the present-day
merger rate.  One parameter left unchanged in past studies, however, was the
\emph{binary fraction} $f_b$, defined as the fraction of stellar
systems that are binaries.  Without loss of generality, our
simulations assume \emph{all} stars form in binaries.   The merger rates
per unit mass implied by a population with a lower binary fraction
$f_b$  can be related to the merger rates we calculate using 
\begin{equation}
{\cal R}(t|f_b) = {\cal R}(t|f_b=1) \frac{f_b(1+\avq)}{1+f_b\avq}
\end{equation}
where $\avq=\left<m_2/m_1\right>$ is the average mass ratio of our initial stellar
population and which is varied between population synthesis models.
In particular, because the effect of this parameter trivially
influences our rate predictions, we can \emph{analytically}
incorporate the influence of any prior assumptions regarding the
binary fraction $f_b$.

The true initial binary fraction of stellar populations is not known;
it is believed to be between $f_{b,min}=15\%$ and $f_{b,max}=100\%$
 \citep{1991AA...248..485D}.  For
this paper we will assume $f_b$ could equally likely take on any value in
this range.
For any given population synthesis model, the binary fraction
and mass ratio distribution multiplicatively influence the observed
detection rate by a factor
\begin{equation}
X\equiv \frac{f_b(1+\avq)}{1+f_b\avq}\; .
\end{equation}
which is distributed between $X_{min}$ and $X=1$ (at $f_b=1$)
according to 
\begin{equation}
\label{eq:ker:Fb:NonLog}
p(X)dX = \frac{1+\avq}{[\avq (X-1)-1]^2}\frac{dX}{1-f_{b,min}}
\end{equation}

\subsection{\usheader: Sampling errors}
\label{sec:smoothing:Sampling}
Roughly speaking, the relatively small  numbers of merging binary black holes
in any simulation 
($m\lesssim 100$) limit our ability to predict the
overall detection rate $\log R_D$ more accurately than $1/\sqrt{m}\ln
10$, because of inaccuracies in reproducing both the mass distribution
$dP/d\mc$ and the delay time distribution $dP/dt$.
Additionally, the relatively small number of \emph{simulations} of
elliptical and spiral galaxies $N_{E,S}\simeq 300-400$ provides little chance of
discovering high-rate models that are more rare than roughly $1/N$.
To allow for these two errors, we must convolve each individual
simulation's merger rate with   kernels reflecting the poisson uncertainty introduced into each measurement by the
  limited number of samples.  For simplicity, we approximate the relative error in $\log N$ as normally distributed with
  standard deviation $1/\sqrt{N}\ln 10$.  Translating this approximation into a PDF, instead of employing a poisson
  posterior distribution we adopt a simple gaussian kernel to describe the relative error in rate given $m$ and $N$
  samples:
\begin{eqnarray}
K_{\rm sim}&=&  K_o(\log R_D, 1/\sqrt{m}\ln 10)\\
K_{\rm samp}&=&  K_o(\log R_D, 1/\sqrt{N}\ln 10)\\
K_o(x,s) &=&  \frac{1}{\sqrt{2\pi s^2}}\exp(-x^2/2s^2)
\end{eqnarray}
As noted above, a third uncertainty is the binary fraction, which we incorporate with the following kernel:
\begin{eqnarray}
\label{eq:def:Kb}
K_b(z, \avq)  &\equiv& \frac{10^z \ln 10}{1-f_{b,min}} \frac{1+\avq}{[\avq (10^z-1)-1]^2} 
\end{eqnarray}
 for $\log 0.15<z<0$.  These terms capture the most significant simulation-related sources of error.
We have also explored and included several other potential sources of error, including (i) calibration uncertainty in the detector
or mismatch in the waveform model (small, typically \orderof{$10\%$})\footnote{Calibration of the amplitude and phase
  relationships between different measured frequencies can significantly perturb results.  The target calibration
  amplitude error
  usually cited is \orderof{$10\%$};  see \cite{LIGO-S3-Calibration} and \cite{LIGO-S1-Calibration} for a detailed survey. }; (ii) uncertainty in the overall star formation history (a factor roughly  $2$) and
in the relative proportions of elliptical and spiral galaxies (roughly $10\%$); and (iii) additional sampling errors introduced
by the sensitivity of the chirp mass average to a few high-mass binaries (\orderof{$0.1/\sqrt{n_{eff}}/\ln 10$}).  
%


\subsection{Model prediction for CBC rate densities }
To review, using the \texttt{StarTrack} population synthesis code, we have
explored a representative sample of stellar evolution produced by
elliptical galaxies ($N_E=\nsimsE$) and spiral galaxies
($N_S=\nsimsS$).\footnote{ We do not   calculate BH-NS or NS-NS merger rates for all $\nsimsE$ elliptical or $ \nsimsS$
    spiral-galaxy simulations; see Table \ref{tab:DataDump} (\onlineOnlyCaveat).  Following the discussion in  the Appendix and \abbrvPSgrbs{}, we only evaluate BH-NS and
    NS-NS merger rates
    when they are adequately resolved ($n$ large) and (to correct for the bias this introduces) adequately unbiased ($n
    N$ large).  Roughly 10\% of   simulations are not used when evaluating the
    NS-NS or BH-NS merger rate distribution.       These models form merging BH-BH binaries much 
    more efficiently than BH-NS or (occasionally) NS-NS binaries.   As discussed in the Appendix, we expect
    observational constraints inevitably rule out the few spiral galaxy models involved. \optional{\editremark{what about ellipticals}
}}
 From these simulations, using the tools indicated in
\S~\ref{sec:popsyn} we carefully extracted their likely properties
(e.g., $dP/dt$), which in turn we convolved with the star formation
history of each of the two major components of the universe
(\S~\ref{sec:sfr}) to generate a preferred present-day merger rate per unit volume ${\cal R}_{k}(0)$ for each model $k$,
for both elliptical and spiral star-forming conditions.

\begin{figure}
\includegraphics{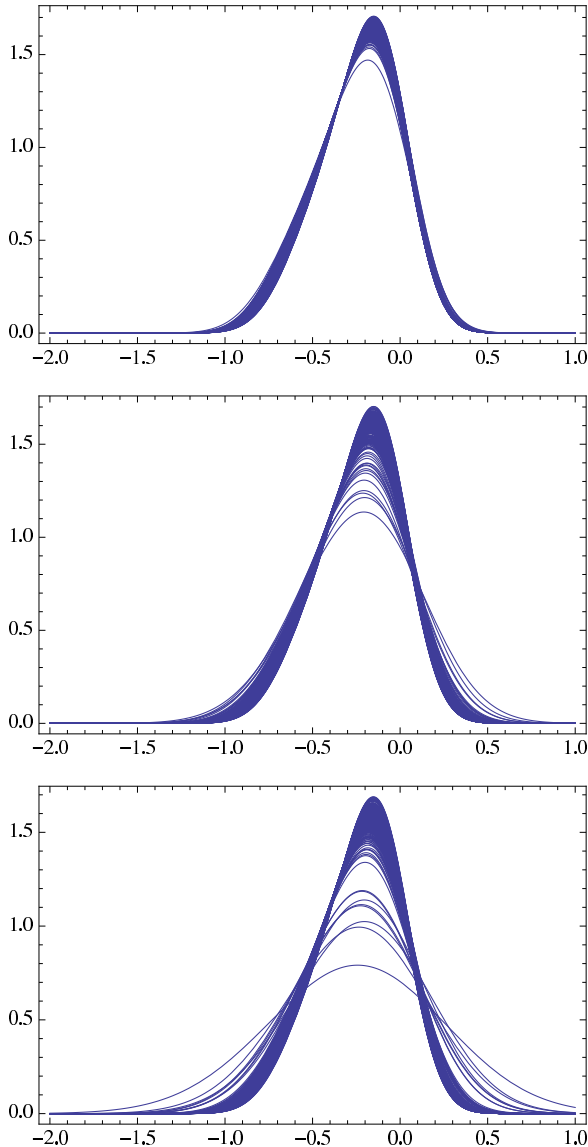}
\caption{\label{fig:Tests:Unconstrained:Smoothing}
All of the kernels $\bar{K}$ versus $\log X$ used in this paper, for NS-NS (top panel), BH-NS (center panel), and BH-BH (bottom panel).
As extremely many binaries are available in each simulation, the NS-NS  simulations' uncertainties are dominated
by the uncertainty in the binary fraction $f_b$.  Because few BH-BH and occasionally BH-NS binaries are available in
each simulation, however, the statistical uncertainty becomes significant.
}
\end{figure}

\begin{figure*}
\includegraphics[width=\textwidth]{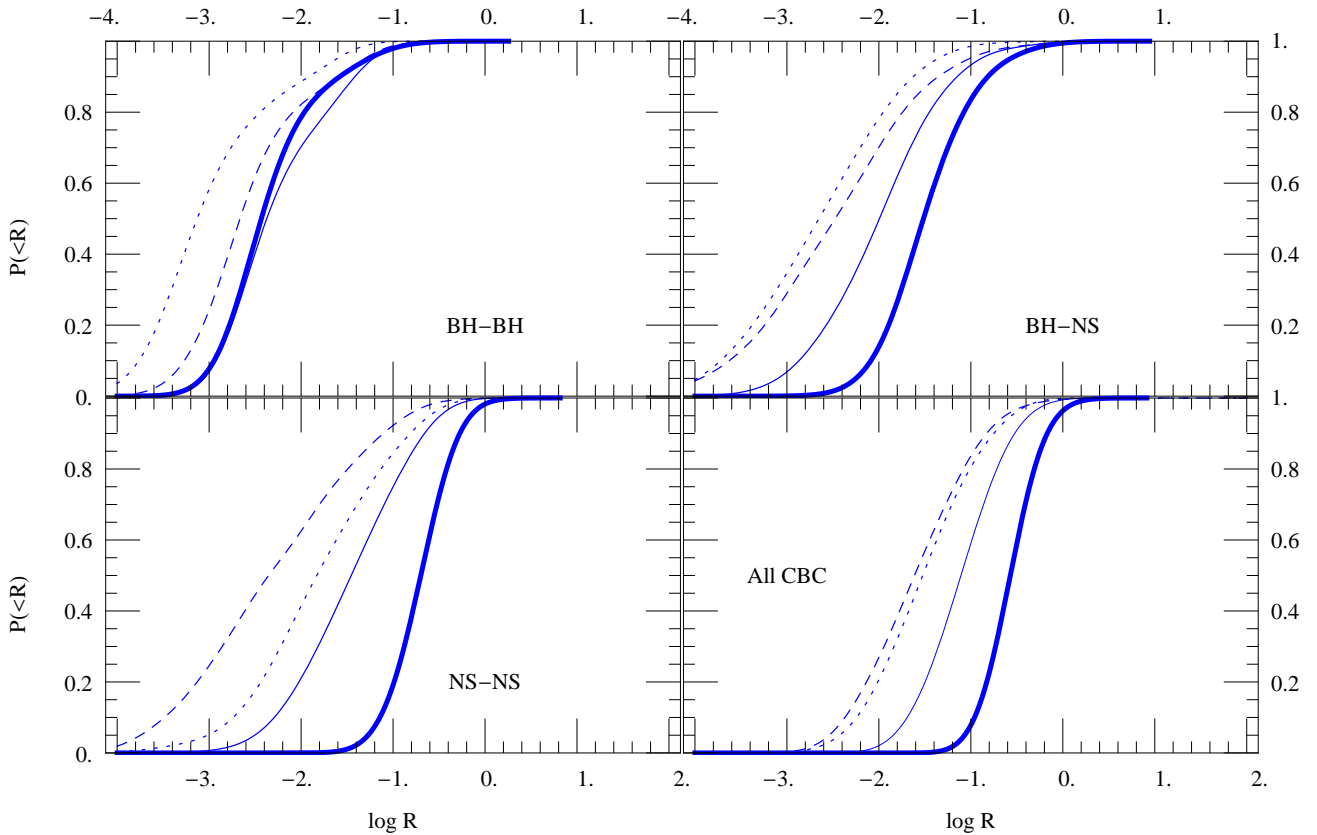}
\caption{\label{fig:volume:rates}Cumulative probability $P(<\log R)$ of various merger
  rates per unit volume (units $\unit{Mpc}^{-3}\unit{Myr}^{-1}$), without (thin) and with (thick) requiring consistency with the binary pulsar population in the Milky Way.  
These distributions incorporate
  uncertainty in the initial binary fraction,  the unknown population synthesis model parameters, and statistical errors implied by the limited size of
    our simulations.   For comparison, the dashed lines show the
  contribution from \emph{ellipticals only}; the dotted lines show the contribution from spiral galaxies \emph{without
    requiring consistency} with the Milky Way.
}
\end{figure*}

Though the set of model universe merger rates $R_{(e,s)k}$ encompasses
many of the most significant modeling uncertainties,  to be more fully
comprehensive and to arrive at a smooth PDF we convolve of the ``model universe'' merger rates
rates $R_{s,p}$ and $R_{e,l}$ for $p=1,\ldots N_S$ and $l=1\ldots
N_E$ with the  uncertainty kernels and binary fraction kernel described  previously
assuming equal prior likelihood for each model (i.e.,
$P(E_l)=1/N_E$ and $P(S_p)=1/N_S$)
\begin{eqnarray}
\label{eq:combine:kernels}
\bar{K} &=&(K_{\rm samp}*K_{\rm sim}*K_b)\\
\label{eq:combine:net:ellipticals}
p_e(\log R_{e}) &=& \sum_{l=1}^{N_E} P(E_l) \bar{K}
   (\log \frac{R_{e}}{ R_{e,l}}) \\
\label{eq:combine:net:spirals}
p_s(\log R_{s})&=&  \sum_{p=1}^{N_S} P(S_p) \bar{K}
   (\log \frac{R_{e}}{ R_{e,p}}) \\
\label{eq:combine:net:add}
p(\log R) &=&  (p_e \oplus p_s) 
\end{eqnarray}
 Figure  \ref{fig:Tests:Unconstrained:Smoothing} shows all of the 
  kernels $\bar{K}$.
In Figure  \ref{fig:volume:rates} the thin ($p(\log R)$) and dashed ($p_e(\log R)$) curves correspond
precisely to the output of this procedure.    The bottom right panel also shows our estimate for the \emph{total}
  merger rate distribution: a three-fold  convolution \footnote{\label{foot:Threefold} Strictly, the merger rate distribution function
      should be calculated by (i) finding the total merger rate predicted by each model (e.g., $R_e =
      R_{bbh}+R_{bhns}+R_{nsns}$); (ii)  centering an error kernel $K$ on that rate, with $n$ set by the total number of
      mergers of all types; and then (iii) convolving the elliptical and spiral distributions together as
      Eq. \ref{eq:combine:net:add}.  
The  approach adopted in the text assumes the joint BH-BH, BH-NS, and NS-NS rate distribution is \emph{uncorrelated}, even though weak
correlations are apparent in Table \ref{tab:DataDump} across our  broad range of parameters
(the table is \onlineOnlyCaveat).   As described in
\S\ref{sec:sub:Discussion:Rates}, this assumption slightly biases us towards higher rates.  On the contrary, the three-fold
     convolution \emph{also} implicitly introduces independent binary fractions for BH-BH,BH-NS, and NS-NS binaries,
     biasing towards lower rates.
Empirically, as can be confirmed using Table \ref{tab:DataDump} this estimate for $p(\log R)$ is in reasonable agreement with the summed
 calculation described above. \optional{\editremark{VERIFY}  } 
Finally, since LIGO detection rates are ubiqutously dominated by the few most massive objects that merge -- i.e., on
$p_{BH-BH}$ -- the total
\emph{detection} rate does not depend sensitively on this order of evaluation issue.
}
\begin{eqnarray}
\label{eq:total:threefold}
p(\log R) = p_{BH-BH}\oplus p_{BH-NS} \oplus p_{NS-NS}
\end{eqnarray}

In the above we assume all simulated population
synthesis models are equally plausible \emph{a priori}.
However, observations of single and binary pulsars significantly
affect our perspective regarding the relative
likelihood of different models.  
NS-NS merger rates derived from the observed sample are in
  fact typically higher than what most population models predict for the Milky Way.
%
We adopt standard  Bayesian
methods to determine the posterior probability $P(S_p|C)$ for a given
spiral-galaxy model $S_p$ ($p=1\ldots N_S$)
\begin{eqnarray}
\label{eq:constraint:SpiralBayes}
P(S_p|C) \propto P(C|S_p)P(S_p)= P(C|S_p)/N_S
\end{eqnarray}
given the observational constraint $C$ on the present-day
spiral-galaxy contribution to the NS-NS merger rate implied by present-day
observations of pulsars.
We adopt as a constraint the requirement developed in \abbrvPSgrbs{} 
from existing NS-pulsar observations, that
the spiral-galaxy merger rate density of double neutron stars lie
between $0.15$ and $5.8 \unit{Myr}^{-1}\unit{Mpc}^{-3}$ (90\% confidence).
The probability $P(C|S_p)$ that the   $p$th spiral-galaxy model predicts a rate consistent with observations of double
neutron stars is found simply by integrating the PDF $P(\log R|S_p)$ for the rate over the constraint 
interval\footnote{%
 Strictly, we should calculate the Bayesian probability with some  $P(\log R|{\cal O})$ that describes the merger rate in
spirals implied by PSR-NS binaries in the Milky Way and uncertainties in the star formation rate density of Milky Way
galaxies.   Our choice of a confidence interval based on broadening the  the  90\% confidence limit for
PSR-NS binaries simplifies the discussion.
}
  $C=[0.15,5.8] \unit{Myr}^{-1}\unit{Mpc}^{-3}$.
After renormalizing the probabilities $P(C|S_P)$ into $P(S_p|C)$ by
requiring $\sum_p P(S_p|C)=1$, we arrive at revised predictions for
\abbrvCBC{} merger rates for BH-BH, BH-NS, and NS-NS binaries as in
Eqs. (\ref{eq:combine:net:ellipticals}-\ref{eq:combine:net:add}) but
with conditional probabilities:
\begin{eqnarray}
p_s(\log R_{s}|C)&=&  \sum_{p=1}^{N_S} P(S_p|C) \bar{K}
   (\log \frac{R_{s}}{ R_{s,p}}) 
\end{eqnarray}
Convolving the constrained spiral rate density distribution $p_s$ with the unconstrained elliptical density $p_e$ leads
to our best estimate for the present-day merger rate per unit volume, shown as the thick solid curve in Figure
\ref{fig:volume:rates}.   Our best results favor merger rates between $2\times 10^{-3}-0.04 \unit{Mpc}^{-3}\unit{Myr}^{-1}$ for
BH-BH mergers (90\% confidence), $0.01-0.28 \unit{Mpc}^{-3}\unit{Myr}^{-1}$ for BH-NS mergers, and $0.11-1.7
\unit{Mpc}^{-3}\unit{Myr}^{-1}$ for NS-NS mergers.

\begin{figure}
\includegraphics{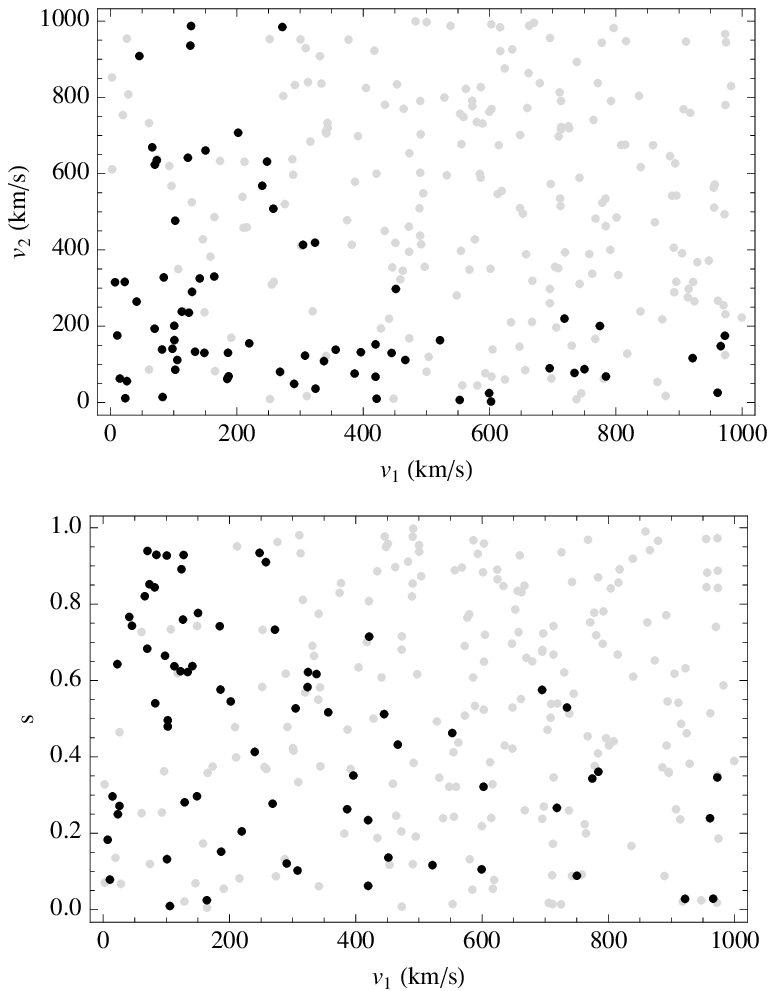}
\caption{\label{fig:constraints:spirals}  Constrained population synthesis parameters for spiral galaxies: compare to the top right and bottom left panels of Figure 5 of
  \abbrvPSmoreconstraints.   In each panel, dark points are included when the corresponding model's posterior probability $P(S_p|C)$ is
  among the top 90\%; gray points are shown for all other $\nsimsS$ spiral galaxy simulations. 
}
\end{figure}

Our previous papers have similarly constrained binary evolution in the Milky Way (\abbrvPSmoreconstraints) and
  in the population of spiral galaxies (\abbrvPSgrbs).   Both interpret observations as  much tighter constraints than
  we do here, because  these papers did not allow for error (e.g., in each
  simulation's prediction and, for the predictions per unit volue, in the star
  formation history of the universe).  Our bayesian constraints are
  significantly less restrictive.    Nonetheless,  as seen in Figure \ref{fig:constraints:spirals} the top spiral
    galaxy models (90\%) have a kick velocity distribution roughly consistent with pulsar observations, similar to that
    recovered in   \abbrvPSmoreconstraints (their Figure 5).
Our predictions for the posterior NS-NS merger rate are conservative for another, technical reason: we implement the
binary fraction as an \emph{uncertainty} rather than a parameter.  We therefore cannot  rule
out  the lowest low binary fractions we allow ($f_b=0.15$) in other spiral galaxies, even though a comparison between binary pulsars in the
Milky Way and our model database disfavors such low values.

\subsection{Discussion}
\label{sec:sub:Discussion:Rates}

\noindent \emph{Two-component universe rarely gives low rates}: The probability we assign to very low rates depends sensitively on our implicit assumption that elliptical
and spiral galaxy star forming conditions are \emph{uncorrelated}.   The convolution  $p_e\oplus
p_s$ [Eq. (\ref{eq:logconvolve})] assigns little probability below the sum of the median elliptical and median
spiral merger rate; see, e.g.,  Figure \ref{fig:volume:rates:ConvolveDemo}.  On the contrary, if all galaxies have the
same undetermined star forming conditions, then the relevant merger rate distribution
\begin{eqnarray}
p(\log R) = \sum_p P(S_p)\bar{K}(\log \frac{R}{R_{e,p}+R_{s,p}})
\end{eqnarray}
assigns more proability to the lowest rates; see, e.g., the bottom panel of Figure \ref{fig:volume:rates:ConvolveDemo}.
Adopting this implicit prior -- identical star forming conditions in elliptical and spiral galaxies --  leads to minimum plausible total BH-BH, BH-NS, and NS-NS merger rates ${\cal R}_{5\%}$ roughly 2.4, 7, and 3.3 times
larger than $\text{max}({\cal R}_{e,5\%}, {\cal R}_{s,5\%})$, where ${\cal R}_{e,5\%}$ is the $5\%$ probability
elliptical-galaxy merger
rate and similarly; compare, e.g., the top
panel of Figure \ref{fig:volume:rates:ConvolveDemo} with Figure \ref{fig:volume:rates}. 

\begin{figure}
\includegraphics{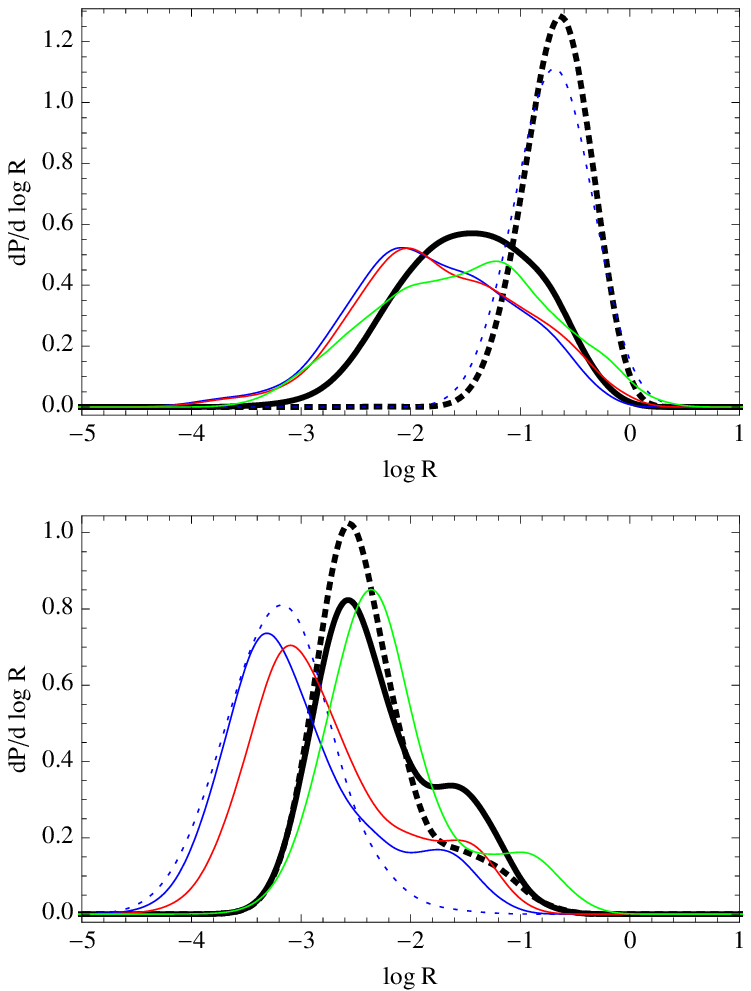}
\caption{\label{fig:volume:rates:Breakdown} Relative likelihood of present-day merger rates $R$ for NS-NS binaries (top panel) and BH-BH binaries (bottom panel)
  given  five assumptions: (1) only nearly steady spiral galaxy star formation produces merging binaries (blue); (2) all galaxies form stars like
spiral galaxies (red); (3) all galaxies form stars like elliptical galaxies (green);  (4) like (1), but using only models that reproduce the present-day Milky Way merger rate (blue, dotted); (5) like
(2), similarly (red, dotted).  For comparison,  the thick black curves shows the fiducial total BH-BH and NS-NS merger
rate (solid curve includes all spiral models; dotted curve only models that reproduce the present-day Milky Way).
}
\end{figure}

\noindent \emph{Ellipticals have different star formation?}: As in  \citet{Regimbau2006-ellipticals}, our fiducial calculations suggest that
elliptical galaxies gave birth to a significant fraction of
all present-day compact object mergers, particularly BH-BH mergers [Figure \ref{fig:volume:rates}].    Our results
inevitably depend on the amount of and conditions for star formation in elliptical versus spiral
galaxies.   On the one hand, we assume most ancient star formation has occurred in ellipticals.  On the other hand, the
Salpeter-like IMFs ($p\simeq -2.3$) adopted for elliptical galaxies produce compact object binaries much more
efficiently than the much steeper high mass slope ($p=-2.7$) adopted for spiral galaxies.\footnote{The steep IMF for
  spiral-like galaxies is motivated by \cite{KroupaClusterAverageIMF2.7}, which theoretically examined the effective
  single-star IMF produced from
  continuous formation of open clusters.  For elliptical galaxies, lacking any definitive suggestion that a similar
  throttled star formation process occurs, we adopt a standard Salpeter IMF.}
To quantify the impact of each factor seperately, in Figure \ref{fig:volume:rates:Breakdown} we compare our fiducial BH-BH and
NS-NS merger rate distributions (thick solid curves) with predictions based on \emph{only} spiral or elliptical star
forming conditions.  Specifically, each new distribution still follows from  Eq. (\ref{eq:int:comovingRate})  [to
get ${\cal
  R}={\cal R}_s$
from our set of spiral-galaxy simulations] and Eq. (\ref{eq:combine:net:spirals}) [to create a PDF, based on ${\cal R}$,
an error kernel, and probability $P(S)$].  
These comparisons allow us to determine the relative impact that three key factors have on our predicted results: (i)
significantly increasing ancient star formation; (ii) requiring ancient star formation produce high mass stars more efficiently than
present-day star formation; and (iii) adopting independent binary evolution models for elliptical and spiral galaxies.
First,  both in distribution and on a model-by-model basis, the total compact
object merger rate due to nearly-steady star formation
(blue curves, which adopt the ``spiral'' star formation rate $\dot\rho=\dot\rho_s$) is less than the amount due to all past star formation history (red
curves, which adopt $\dot{\rho}=\dot{\rho}_e+\dot{\rho}_s$) only by a simulation-dependent factor between 
$1.15$-$1.3$ for NS-NS and between $1.5$-$2$  for BH-BH binaries.\footnote{The merger rates per unit volume are higher than an extrapolation of the Milky Way's rate to a volume density
    by a factor of roughly $[\int dP_{m,NS-NS}/dt \dot{\rho}(t)]/\dot{\rho}(0)P_{m,NS-NS}(<13 \unit{Myr})$.
}
%
%
Second, on a model-by-model\footnote{Though not used or provided here, we have compared several pairs of binary evolution models which
  differ only in their IMF and metallicity.} and distribution basis, provided the same amount of star formation our
elliptical galaxy models produce merging binaries roughly 
$1$ (NS-NS) or $1-10$ (BH-BH) times more frequently than spiral-only star forming conditions (both multiplied by factors
of order unity),
entirely due to their shallower IMF and broader metallicity distribution; compare the red and green curves in Figure 
\ref{fig:volume:rates:Breakdown}.
Third, though ellipticals produce merging binaries more efficiently per unit mass, fairly few BH-NS and NS-NS binaries
remain to merge after such a long delay ($dP/dt\propto 1/t$).  For BH-NS and NS-NS binaries, elliptical galaxies'
present-day merger rate balances two positive factors -- very efficient compact binary formation (e.g., $\sim 3$
times higher, based on the IMFs) 
combined with a high rate of early star formation  ($\sim 10$ times higher)  -- against an $\sim 30$
lower merger rate $dP/dt$ via a few Gyr vs $\sim 100 Myr$ typical delay; see the discussion in \abbrvPSgrbs.    For BH-NS and NS-NS binaries, this balance roughly leads to
comparable merger rates in elliptical and spiral galaxies.\footnote{In fact, the spiral NS-NS merger rate is
  significantly higher than the elliptical NS-NS rate, due to the
existence of a population of ultracompact binaries not well described by this argument.}  
%
To summarize, our fiducial unconstrained predictions differ from previous predictions based on steady, spiral-galaxy star formation  by three factors: (i)
the addition of ancient star formation, increasing the overall merger rate above steady spiral-only by a factor
$1.15-1.3$ for NS-NS and $1.5-2$ for BH-BH; (ii) the conversion of ancient star
forming conditions from spiral-like to elliptical, increasing the total \emph{present-day} merger rate by a factor of
order unity for BH-NS and
NS-NS binaries; and
 (iii) the assumption that elliptical and spiral star forming conditions are independent, biasing against low total
 merger rates.\footnote{ Additionally, we apply observational constraints only to spiral galaxies, insuring the retension
   of elliptical galaxy
 models with the highest BH-BH formation rates.  As discussed in  \S\ref{sec:constraints}, the spiral-galaxy models
 with frequent-enough NS-NS mergers generally have $\alpha \lambda$ not too small.  Very inefficient common envelope
 evolution is required to produce the highest  BH-BH merger rates.
}
%
%

\begin{figure}
\includegraphics[width=\columnwidth]{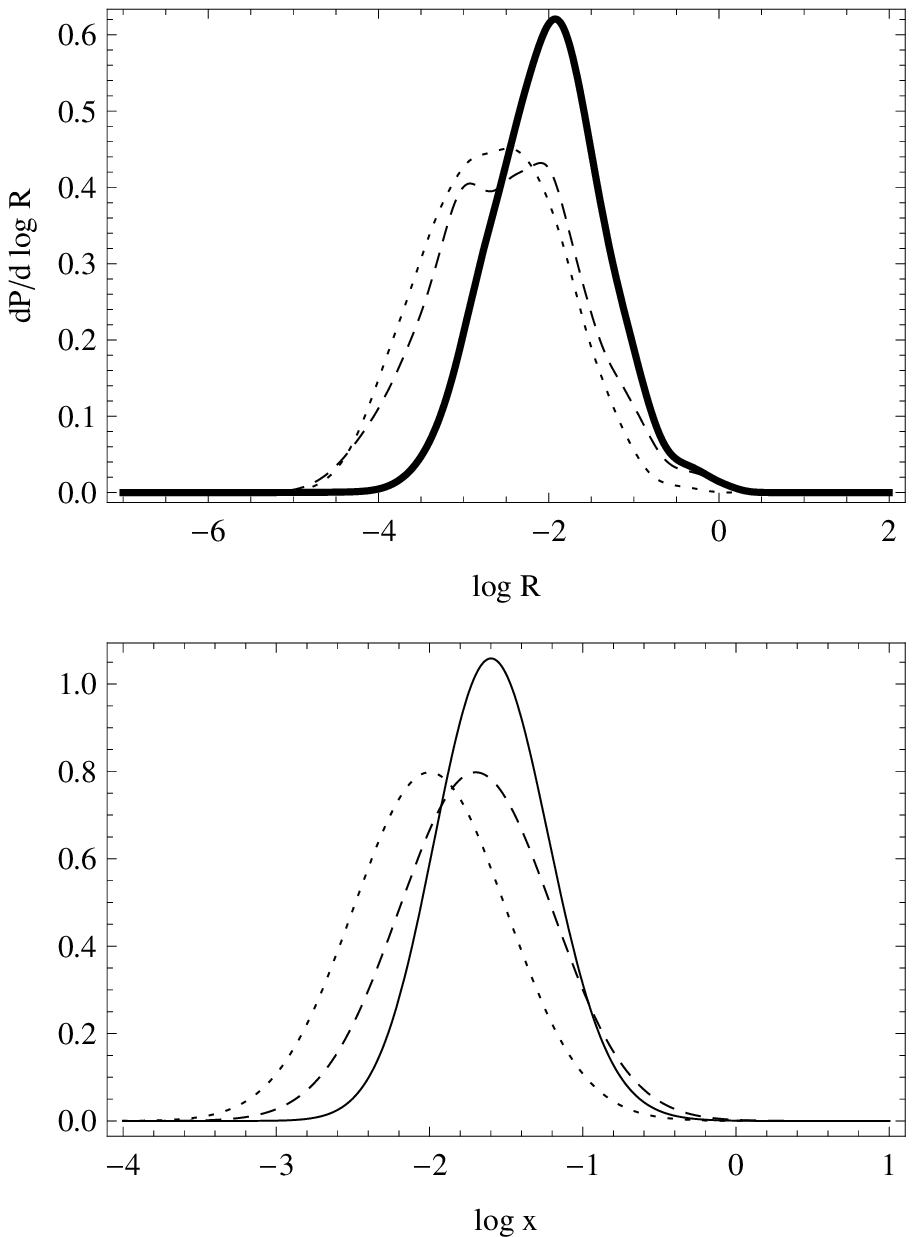}
\caption{\label{fig:volume:rates:ConvolveDemo}Convolutions illustrated.
Top panel: the BH-NS merger rates ${\cal R}$ due to ellipticals (dashed), spirals (dotted), and
  overall (solid) assuming uncorrelated elliptical and spiral star forming conditions, as in the text and Figure \ref{fig:volume:rates}.  The minimum likely merger
  rate overall is significantly larger than the minimum likely due to ellipticals or spirals alone.
Bottom panel: a gaussian distribution $p(\log x)d\log x$ (dotted); $p\oplus p$ (solid), analagous to assuming a
two-component universe where elliptical and spiral galaxy rates are independently drawn from $p$; and the distribution
for $y=x+x$ (dashed), analagous to a two-component universe with identical elliptical and spiral star forming
conditions, both drawn from $p$.  
}
\end{figure}

\noindent \emph{Ellipticals ``dominate'' BH-BH rate}:  Adopting the same binary evolution model  but
different star formation histories,  elliptical galaxies usually produce more
merging BH-BH binaries at present than spiral galaxies.   In this sense, the elliptical galaxy BH-BH merger and
detection rates  ``dominate'' the total merger rate.  However, some elliptical galaxy models do yield  lower merger rates
than some spiral galaxy models; neither \emph{ubiqutously} dominates.  For example,  Figure
\ref{fig:volume:rates:ConvolveDemo} suggests that low but plausible total  BH-NS
merger rates require more mergers in spirals than ellipticals; high but plausible merger rates require the opposite.

\noindent \emph{Binary evolution priors}:   Our predictions depend  on the relative likelihood of different
binary evolution model parameters.  To facilitate comparison with \abbrvPSmoreconstraints{} and to avoid omitting any vital region
of parameter space,   we allow and treat as equally plausible a
 wide parameter range,  including extremely large (and empirically implausible)  supernova
kick magnitudes $v_{kick}\simeq 1000\unit{km/s}$.   
The reader can reproduce our results or explore alternative  assumptions by reweighting the data provided in Table
\ref{tab:DataDump}, \onlineOnlyCaveat.

\noindent \emph{Binary evolution parameters}: Finally, as outlined in \S~\ref{sec:sub:ModelUncertaintyReview}, we fix a
number of assumptions which can dramatically influence merger and even detection rates.
For example,  the maximum NS mass $m_{+NS}$ completely determines the nature of  final compact binaries, as  it subdivides a roughly
$m_{+NS}$-independent mass distribution into three regions: NS-NS, BH-NS, and BH-BH.   Evidently the
individual \emph{component} merger rates depend sensitively on $m_{+NS}$.
  However,
as relevant gravitational wave interferometers  cannot easily distinguish between two compact objects of comparable mass $1
M_\odot <m<3 M_\odot$, the \emph{overall detection rate} described later depends only  weakly on $m_{+NS}$.
Similarly, while the neutron star birth mass distribution can influence the relative proportions of each type, as well as
the observable pulsar binary mass distribution, our results should not depend significantly on the amount distribution
of masses.  

Several other binary evolution parameters not explored here have previously been shown to influence merger rates
significantly, albeit much less significantly than the parameters we explored here.
First, as noted previously in \S \ref{sec:sub:ModelUncertaintyReview}, we adopt the full Bondi-Hoyle accretion rate onto black
holes in common envelope, rather than the smaller rates suggested by recent hydrodynamical simulations;
cf. \cite{ChrisBH2007,ChrisSpinup2007}.   A smaller mass accretion rate, if adopted, would reduce the incidence of
accretion induced collapse, marginally changing the proportions of BH-BH, BH-NS, and NS-NS binaries; marginally lower the masses of some merging binaries,
increasing the  factor by which common envelope evolution tightens orbits; and generally change merger and detection rates by a small factor compared with the
overall range considered  [Footnote \ref{foot:AlternateBondi}].
Second, we examined only  isotropic kick distributions.  Several authors have suggested spin-kick alignment on
theoretical and empirical grounds; see,e.g., \cite{2008AIPC..983..433K}, \cite{ns-polarkicks-rateimplications-Kuranov2009} \cite{2007ApJ...656..399W} and references therein.   As the same mass transfer that brings binaries
close enough to merge also aligns their spins and orbit before the second supernova, polar kicks are particularly
efficient at disrupting binaries; see   \cite{ns-polarkicks-rateimplications-Postnov2007}.
Though polar kicks would dramatically transform our predictions, the observed PSR-NS merger rate in the Milky Way
not only corresponds to a \emph{high}, not low rate\footnote{While the range of predicted Milky Way NS-NS merger rates depends
  on the adopted IMF, even a standard Salpeter IMF leads to a range of predictions that intersects PSR-NS observations
  only for high rates.}  but also implies model supernova kick magnitudes comparable to observed
pulsar proper motion velocities [\abbrvPSmoreconstraints{}, as well as  our Figure \ref{fig:constraints:spirals}].
Third, we adopt a specific model for mass loss in massive stars; by changing the evolution of massive progenitors,
alternate stellar wind models can dramatically modify merger rates (cf. \cite{popsyn-LowMetallicityImpact-Chris2008}).
Fourth, we adopt a single, time-independent  metallicity for each type of star forming region.   Binary merger rates
can depend sensitively on metallicity, particularly through metallicity-dependent stellar winds \citep{popsyn-LowMetallicityImpact-Chris2008}.
A future publication will explore the latter two issues in more detail.

\noindent \emph{Advanced detectors and redshift-dependent mass evolution}:  For simplicity we have adopted
a single  time-independent  mass distributions for each type of merging binary: BH-BH, BH-NS, and NS-NS.  Looking
backward to higher redshift, however, the 
relative proportions of each binary type change, as each possesses a different delay time distribution $dP/dt$.   In
other words, despite using  ``time-independent'' mass distributions,
the \emph{instantaneous total mass distribution}, found by gluing these three distributions  together in proportion to
their instantaneous merger rates ${\cal R}$, will vary with time.    Networks of advanced detectors and particularly third generation
detectors will probe a time-dependent mass distribution.  The authors and collaborators will also address this
issue in a future paper.


\section{\abbrvCBC{} detection rates and detection probability}
\label{sec:popsyn:nonbh}






\subsection{\abbrvCBC{} Detection Rates for LIGO }
\label{sec:detectionRates}

\begin{figure*}
\includegraphics{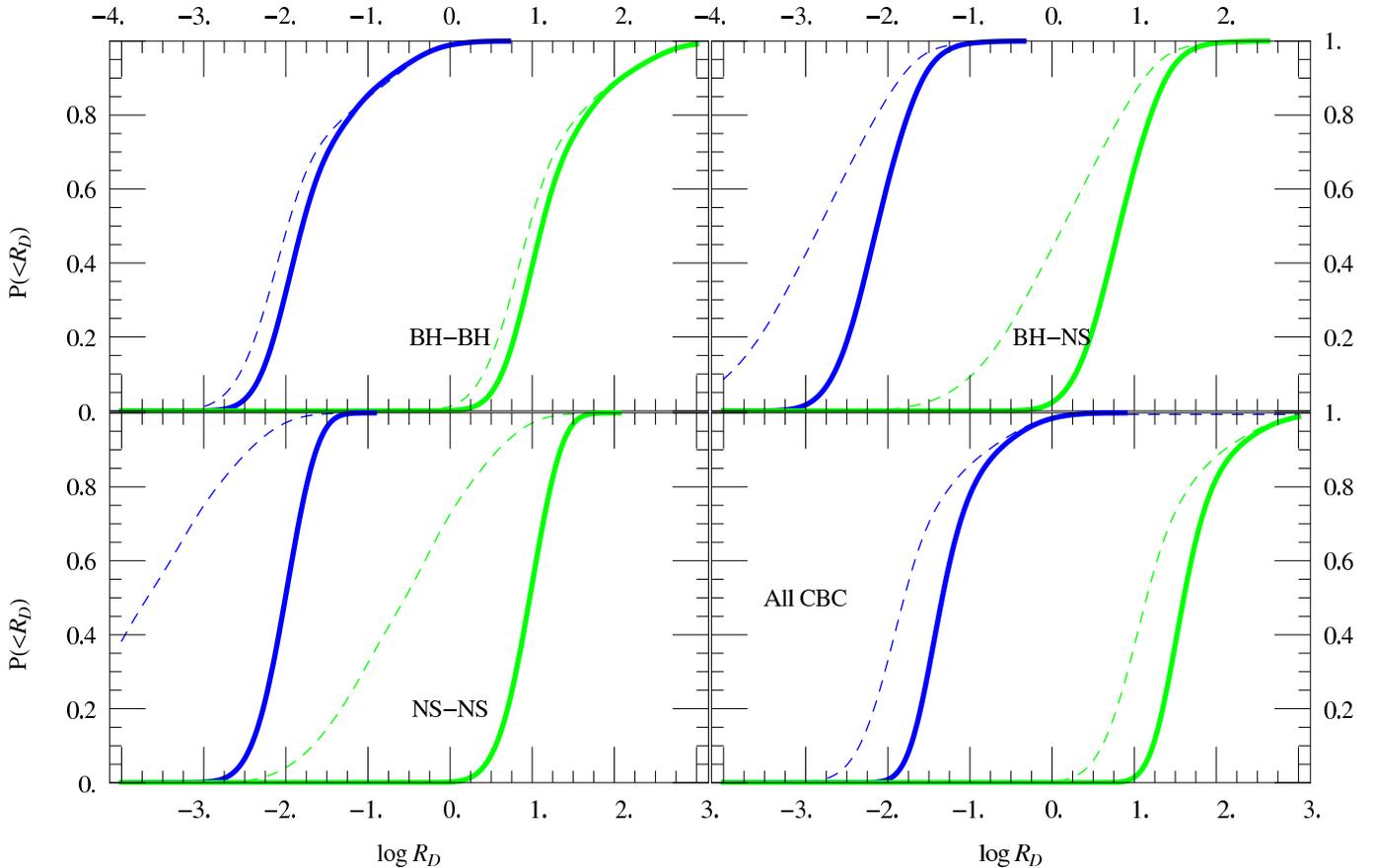}
\caption{\label{fig:ligo:rates} 
Cumulative probability $P(<\log R_D)$ of various event
  rates for  initial (blue; assuming $\Dh =31\unit{Mpc}$) 
and advanced (green; assuming $\Dh=445\unit{Mpc}$) single LIGO
  interferometers  as part of a network,  requiring consistency with the binary pulsar population in the Milky Way.  
These distributions incorporate
  uncertainty in the initial binary fraction, as discussed in the
  text, as well as uncertainty in the many population synthesis model
  parameters.  For comparison the dashed curves show the 
contribution from \emph{ellipticals only}. 
}
\end{figure*}

Using the \texttt{StarTrack} population synthesis code, we have
explored a representative sample of stellar evolution produced by
elliptical galaxies ($N_E=\nsimsE$) and spiral galaxies
($N_S=\nsimsS$).
 From these simulations, using the tools indicated in
\S~\ref{sec:popsyn} we carefully extracted their likely properties
(e.g., $dP/dt$), which in turn we convolved with the star formation
history of each of the two major components of the universe
(\S~\ref{sec:sfr}).
Using our understanding of the LIGO range
(\S~\ref{sec:ligo}) we can convert these merger rate histories into
expected  single-interferometer LIGO detection rates.
Finally, though the set of model universe detection rates $R_{D,(e,s)k}$ encompasses
many of the most significant modeling uncertainties,  to be more fully
comprehensive and to arrive at a smooth PDF we plot in Figure
\ref{fig:ligo:rates} the convolution of the ``model universe'' detection
rates $R_{D,s,p}$ and $R_{D,e,l}$ for $p=1,\ldots N_p$ and $l=1\ldots
N_E$ with the  uncertainty kernels and binary fraction kernel described  previously
assuming equal prior likelihood for each model (i.e.,
$P(E_l)=1/N_E$ and $P(S_p)=1/N_S$)
\begin{eqnarray}
\bar{K} &=&(K_{\rm samp}*K_{\rm sim}*K_b)\\
\label{eq:combine:net:ellipticals:Constrained}
p_e(\log R_{D,e}) &=& \sum_{l=1}^{N_E} P(E_l) \bar{K}
   (\log \frac{R_{D,e}}{ R_{D,e,l}}) \\
\label{eq:combine:net:spirals:Constrained}
p_s(\log R_{D,s})&=&  \sum_{p=1}^{N_S} P(S_p) \bar{K}
   (\log \frac{R_{D,e}}{ R_{D,e,p}}) \\
\label{eq:combine:net:add:Constrained}
p(\log R_D) &=&  (p_e \oplus p_s) 
\end{eqnarray}
and subsequently combine the elliptical and spiral detection rate
distributions [Eq. (\ref{eq:logconvolve})].
[Most of these convolutions can be performed trivially by adding the
standard deviations of gaussians in quadrature; the remaining two
convolutions are performed numerically.]     As previously we estimate the total detection rate via a three-fold
  convolution [Eq. \ref{eq:total:threefold}].\footnote{ Per the discussion of Footnote \ref{foot:Threefold}, this
      convolution only approximates the true detection rate.  In practice, one factor usually dominates the detection
      rate in each component (e.g., for ellipticals, BH-BH mergers).   
}
Though the same basic techniques applied above can be and have been applied
to BH-NS and NS-NS binaries previously  (\abbrvPSgrbs),
we now
incorporate error propagation, a simulation-by-simulation estimate of the relevant chirp mass, and a variable binary fraction; the merger rate
distributions they provide therefore are not proportional to our predictions.


Figure
\ref{fig:ligo:rates} shows the distribution of \abbrvCBC{} detection rates predicted by
Eq. (\ref{eq:def:DetectionRate:Local}) for  individual initial  LIGO interferometers as part of a network (thin curves) as well as the contribution from
elliptical galaxies alone (dashed curves).\footnote{ The raw and postprocessed data
used to create these figures is available from the first author on request.}
 As indicated by the close proximity of the elliptical and total merger rate
distributions, most BH-BH  mergers that LIGO detects should be produced in elliptical galaxies.   This
elliptical bias arises because  BH-BH binaries have long delays between their formation and
eventual merger; their present-day merger rate is more easily influenced by ancient star formation. 
On the other hand, the difference between the dashed and thin curves in the bottom left panel indicate that the NS-NS
merger rate must be \emph{spiral}-dominated.


Because the initial  LIGO detectors essentially probe different-sized volumes of the local universe for each merger type, their
detection rate distributions are  identical to those of any similar-scale single- or multiple-IFO network,  mod a
  constant horizontal offset determined by the relative increase in volume NS-NS binaries can be seen [$\log V_{NS-NS}/4\pi (14 \unit{Mpc})^3/3$].  Similarly,
because a single  advanced LIGO detector's reach to  NS-NS and BH-NS binaries is often not cosmologically significant,
weighting over detector and source orientation, our predictions for NS-NS and BH-NS detection rates are identical to
those for initial LIGO, mod an offset [$=\log (197/14)^3=3.44$].  However, the typical range of a single detector to BH-BH binaries
\emph{does} extend past $z\simeq 0.25$, where the star formation rate is noticably higher and where cosmological
  volume and redshift factors become significant.  But because of the long
delay between BH-BH merger and progenitor birth, the BH-BH merger rate $R(t)$ generally increases much more slowly with
redshift than the star formation history.  As a result, the single-IFO BH-BH detection rate distribution is also
comparable  to but slightly less than
what we expect after scaling up initial-LIGO results to larger range: a detailed calculation suggests
a merger detection rate between $3 -300\unit{yr}^{-1}$ (90\% confidence), in good agreement with the $20-500
\unit{yr}^{-1} (\Cv/197\unit{Mpc})^3$ (90\% confidence) expected from cubic rescaling of the initial LIGO 
result.\footnote{  In our simulations, the  black hole-black hole rates  generally are \emph{constant} or \emph{decrease} with redshift, due to
  the long delay between binary birth and merger.
Also, the ``redshifting factor'' $1+z$ in the detection rate integral implies a slightly smaller comoving 4-volume swept out by
the detector's past light cone per unit time at the maximum redshift:  $\int dz dV_c/dz/(1+z)\simeq V_c/(1+z)$.  Both
these factors suggest the BH-BH detection rate at high redshift will be \emph{lower} than a simple cubic extrapolation
of the local-universe result.
}
%
For networks of advanced detectors, however, cubic scaling breaks down severely; the range is significant enough to require a full cosmological treatment
and network-sensitivity-beampattern averaging.

\subsection{Comparisons with pulsar   observations of Galactic NS-NS binaries}
\label{sec:constraints}

\optional{
\begin{figure}
\includegraphics[width=\columnwidth]{fig-Unconstrained-MergerRateDistributions-Modified}
\caption{\label{fig:results:mergerRates} 
Distribution of the
  contribution of spiral galaxies (blue) to the present-day  NS-NS merger rates expected \emph{a priori}
  (dotted) and after requiring consistency with the pulsar population in the Milky Way (solid).   For
  comparison,  the \emph{a priori} (black, dotted) and constrained (black, thick) predictions for the total NS-NS rate
  are also shown.    The solid
  gray region on the right indicates the interval of NS-NS merger
  rates per unit volume $C=[0.15,5.8] \unit{Myr}^{-1}\unit{Mpc}^{-3}$ (\abbrvPSgrbs, based on \cite{Chunglee-nsns-1}) that are consistent with the present-day Milky
  Way population and the density of Milky Way-like galaxies in the
  universe.
}
\end{figure}
}

Following the previously-discussed procedure to re-evaluate the likelihood of our
spiral-galaxy population synthesis models, we arrive at a distribution
of initial LIGO detection rates as indicated in Figure 
\ref{fig:ligo:rates}.   Comparing these distributions with
the unconstrained results of the previous calculation, we arrive at
the following conclusions: 

\noindent \emph{Higher NS-NS detection rate}: As seen in Figure
\ref{fig:volume:rates}, inferences drawn from the double pulsar
population about the  present-day NS-NS merger rate lie on the high
end of what our models can presently reproduce.   By requiring our
models reproduce those observations, our constrained distributions
naturally favor higher NS-NS detection rates.

\noindent \emph{Lower BH-BH detection rate from spirals}:
Compared to our previous results (Fig. \ref{fig:ligo:rates}),  the highest
binary black hole merger and thus detection rates are ruled out
\emph{in spiral galaxies}.    Physically, the observational constraint supports those population synthesis models where
one particular parameter ($\alpha \lambda$, related to common-envelope evolution) is not exceptionally small ($\alpha
\lambda>0.1$); see for comparison the top left panel of 
Figure 5 in \abbrvPSmoreconstraints.  In turn,  this parameter  $\alpha \lambda$ correlates strongly with the BH-BH
merger rate, with  low values of $\alpha \lambda$ being required to produce the largest BH-BH merger rates.  Thus,
because observations of NS-NS binaries in
the Milky 
Way favor high NS-NS merger rates, they disfavor the highest rates of BH-BH mergers in spiral galaxies.

\noindent \emph{Observations of the NS-NS merger rate in spiral
  galaxies do not 
  constrain ellipticals, BH-NS well}:
With the
exception of NS-NS mergers, Figure \ref{fig:volume:rates} demonstrates that  most \emph{constrained} predictions for
 merger rates (and, though not shown here,  LIGO's detection rate) agree strikingly well  with our \emph{a priori} 
expectations.   These consistent predictions should be expected, since
observations of spiral galaxies cannot constrain differences (e.g.,
due to metallicity and IMF) in the binary evolution in elliptical
galaxies.  In particular, the single most likely source of a
gravitational wave detection in the near future, BH-BH mergers, are
due to their long characteristic ages expected to occur primarily in 
 in elliptical galaxies; as a result, the present-day BH-BH and total
 \abbrvCBC{} detection rate depends only weakly on our ability to rule
 out certain population synthesis parameters in spiral galaxies.
Furthermore, even in spiral galaxies alone, as seen in \abbrvPSmoreconstraints{}  and references
therein, the NS-NS and other \abbrvCBC{} merger rates are not tightly
correlated.  Thus despite the fairly strong and consistent constraints
that double pulsar observations imply for \emph{binary evolution
  parameters}, as described in detail in \abbrvPSmoreconstraints, our overall expectations for LIGO's detection rate remain
largely unchanged.

\subsection{Net probability of \abbrvCBC{} detection}

\begin{figure}
\includegraphics[width=\columnwidth]{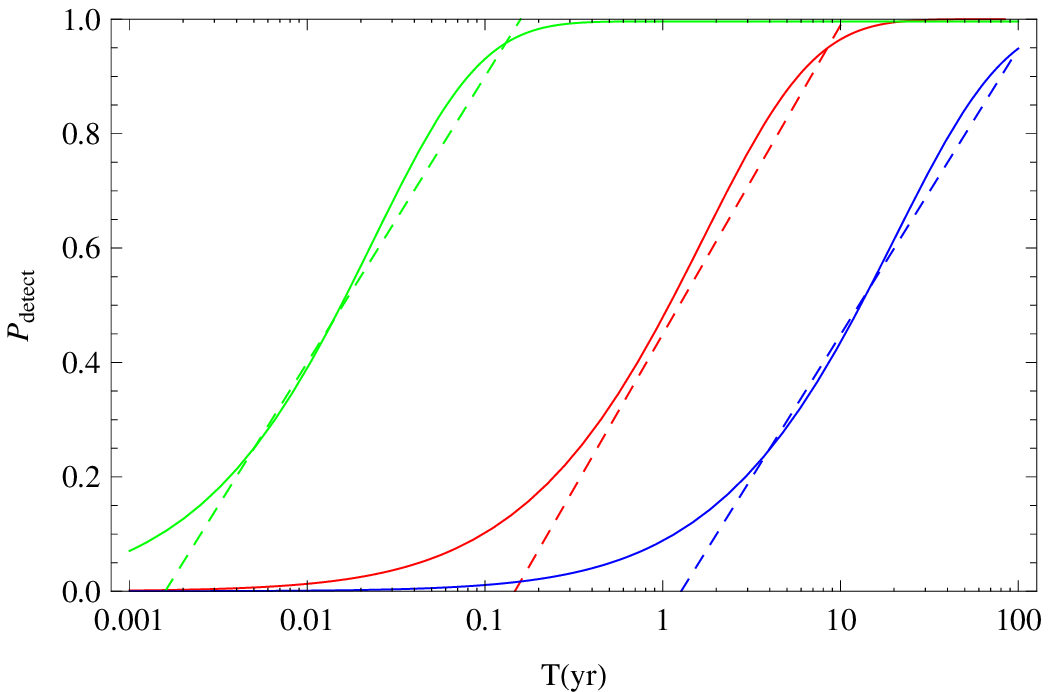}
\caption{\label{fig:results:priorDetectionProbability} 
Probability
  $P_{\text{detect}}(\ge 1|\Cv,T)$  
  [Eq. (\ref{eq:def:pdetect})] of
  detecting at least one \abbrvCBC{} merger in time $T \unit{yr}$
  by a single interferometer  as part of a network with volume-averaged range $\Cv=14 \unit{Mpc}$ (solid blue), $32 \unit{Mpc}$ (solid red) and $197\unit{Mpc}$ (solid green) to
  double-neutron star inspiral,  taking into account information about binary pulsars
  in the Milky Way.  For comparison, the dashed curves
corresponds to the approximations presented in Eq. 
\ref{eq:results:constrained:pdetect}. 
}
\end{figure}

Factoring in poisson statistics for each model, the \emph{net} prior
probability $P_{d}$ that a single LIGO interferometer with a
single-IFO  range to NS-NS inspiral of  $\Cv$  operating for time $T$ 
can make at least one detection is
\begin{eqnarray}
\label{eq:def:pdetect}
P_{\text{detect}}\left(\ge 1| \Cv,T \right)&=& 
 \int dR_D p(R_D)(1- e^{-R_D   T}) 
\end{eqnarray}
where $\Cv$ implicitly enters as a scale factor for $R_D$ ($\propto \Dv^3$).
More generally, for a multi-interferometer configuration with a time-varying network sensitivity,
the appropriate detection probability can be easily calculated from our reference probability by replacing $T$ in the
above expression by $\int dt (\Cv(t)/\Cv_{ref})^3$; the latter quantity characterizes the 4-volume to which the
search is sensitive.
This detection probability can be
significantly increased by  improvements in the  LIGO
detector or run schedule that improve the detection 4-volume $V T$. 
 Assuming a fixed, orientation-averaged range  $\Cv$  for any fraction of the whole LIGO network,  the total
probability of detecting one or more events by that fractional network  can be approximated by 
\begin{subequations}
\label{eq:results:constrained:pdetect}
\begin{eqnarray}
P_{\text{detect}}(\ge 1|14\unit{Mpc},T)&\simeq& 0.4 + 0.5 \log (T/8 \unit{ yr})  \\
P_{\text{detect}}(\ge 1|197 \unit{Mpc},T)&\simeq& 0.4 + 0.5 \log (T/0.01\unit{ yr}) 
\end{eqnarray}
\end{subequations}
for $T$ between $1 \unit{yr}$ and $100\unit{yr}$ or $2\times 10^{-3}$ and $0.1\unit{yr}$, respectively.  
These approximations are shown in Figure
\ref{fig:results:priorDetectionProbability}.  

Equivalently,  our calculations suggest that  a gravitational-wave network must have a 4-volume sensitivity of  $1.3\times
10^{5} \unit{Mpc}^3\unit{yr}$ (roughly
equivalent to a $32 \unit{Mpc}$ sphere for 1 year) in order
to have  a  50\% chance of detecting one or more events, as
\begin{eqnarray}
P_{\rm detect}(\ge 1|\Cv, 1\unit{yr})\simeq 0.5 + 1.5 \log \Cv/32\unit{Mpc}
\end{eqnarray}
for $\Cv$ between 20 and 70 Mpc.

\begin{table}
\caption{\label{tab:DataDump}Parameters of all binary evolution simulations used in this paper, along with predicted  merger rates $R$,
  detection rates $R_D$, and (for spiral simulations) posterior probabilities.  This table is 
\onlineOnlyCaveat.}
\end{table}

\optional{
\subsection{Discussion(*)}


* IMF issue : does it influence detection rates over and above, by increasing number of high-mass BHs?

\editremark{part 2: comparison to previous results}: I need a model-by-model comparison.  We have a huge table of model
results; \editremark{CITE IT}.  Draw some fiducial comparisons to previous popsyn studies, where available (e.g., some
earlier StarTrack results, which do certain parameter slices)...mention not a grid, so tricky to do.
}

\subsection{Comparison to earlier studies}

%
Our fiducial predictions for compact object merger rates [Fig. \ref{fig:volume:rates}] and
gravitational wave detection rates 
[Fig. \ref{fig:ligo:rates}] reflect our uncertainties about the star-formation conditions and history of the universe.
For example, results consistent with current observational constraints  require compact binary formation in elliptical galaxies  and therefore disallow the worst-case
scenario: mergers formed only in spirals by our least-efficient spiral-galaxy model.
Nonetheless, our broad parameter study encompasses many scenarios and predictions.  For $100\%$ binary fraction and for each type of compact binary (NS-NS, BH-NS, and BH-BH),  the most
optimistic and pessimistic single-component scenarios in our comprehensive table of results (Table \ref{tab:DataDump},
\onlineOnlyCaveat)\footnote{For each simulation used in this study, this table provides input parameters;  the number of
  each type of binary appearing $10\unit{Gyr}$ after a starburst ($n$ in the appendix); the number of binaries predicted
  to merge within the age of the universe ($m$ in the appendix); and, after convolving with a suitable star formation
  rate, the present-day merger and gravitational-wave detection rate for each binary type.
} correspond to
merger rates extending between
 \orderof{ $10^{-4}-1$  $\unit{Myr}^{-1}\unit{Mpc}^{-3}$}.

Detailed quantitative comparisons of our  results to others appearing in the literature must  account for alternate
normalization and star formation history choices, which often vary significantly from study to study.
Our constant star-formation spiral-galaxy predictions\footnote{Most predictions in the literature
  present results only for the Milky Way, based on an implicit or explicit steady star formation rate assumption.  These results can be translated 
  using our present-day spiral-galaxy star formation rate (\S \ref{sec:sfr}) and the density of Milky Way-equivalent
  galaxies, by blue luminosity ($\simeq 0.01 \unit{Mpc}^{-3}$); see, e.g., \cite{LIGO-Inspiral-s3s4-Galaxies}.
} in Table \ref{tab:DataDump} are comparable to Table 4 in the review article by \cite{lrr-2006-6},  to Table I in
\cite{StarTrack}, to Tables 2 and 3 in \cite{ChrisBH2007}, and to Table 5 in \cite{2002MNRAS.329..897H}.
Our 
constant star-formation  elliptical-galaxy\footnote{These papers adopt a Salpeter-like IMF but steady
  star-formation rate, a combination not provided in our Table \ref{tab:DataDump}.  For a steady star formation rate,
  the merger rate can be estimated from two entries in the table: the ratio of the number of merging binaries ($m$) to
  the star formation time ($T$), corrected for the mass ratio distribution; see the Appendix of \cite{PSconstraints}.}
predictions can be compared to
Tables 4, 6, and 7 of  \cite{ADM:Vos2003} and to Table 1 in \cite{1998AA...332..173P}.
Our results assuming all galaxies have spiral-like star forming conditions [cf. Fig \ref{fig:volume:rates:Breakdown}]
can be compared with \cite{2003ApJ...589L..37B}. 
Finally, our constant and time-dependent elliptical galaxy results are directly comparable to \cite{Regimbau2006-ellipticals}.
Even after allowing for different inputs, a model-by-model comparison of our results with others available in the literature illustrate physical differences between different authors'
population models and our own [\S~\ref{sec:sub:ModelUncertaintyReview}].  
 For example, unlike \cite{ADM:Vos2003}, rather than calculate the envelope binding energy fraction
$\lambda$,  we fix a single factor $\alpha \lambda$ for all binaries in
common envelope, leading to smaller BH-BH merger rates for comparable parameters; see, e.g., the discussion in \cite{lrr-2006-6}.
As discussed in \S \ref{sec:sub:Discussion:Rates}, unlike \cite{ns-polarkicks-rateimplications-Postnov2007}  we adopt isotropic kicks, avoiding the least favorable kick
configurations and merger rates; see their Figure 3.  Unlike model C in \cite{ChrisBH2007}, we assumed that the common-envelope evolution of a Hertzsprung gap donor led to stellar merger, not a
compact binary.   We apply a similar kick to all NS, rather than assign a different kick based on its prior (binary)
evolution history \citep{2002ApJ...574..364P} or its supernova mechanism (i.e., a low kick for electron-capture
supernova; see 
\cite{2009ApJ...699.1573L}, \cite{2008AIPC..983..433K} and references therein).
Additionally, as described in \S~\ref{sec:sub:ModelUncertaintyReview} and \S~\ref{sec:sub:Discussion:Rates}, we make several other specific choices consistent
with our previous work but not universally adopted in the literature (e.g., for the
Bondi accretion rate in episodes of hypercritical common envelope accretion; for maximum  NS mass).  
 However, a detailed exploration of all these correlated effects would require coordinated simulations performed for the
purpose of detailed comparisons; such an undertaking is beyond the scope of this paper.  Instead, we are satisfied that
our results are broadly consistent with early studies of the Milky Way, particularly given the number and parameter
survey breadth of the model database used here compared with earlier studies by other groups.  For comparison, in the
analysis of our present results we do make detailed comparisons to earlier studies from our group (\S~\ref{sec:sub:Discussion:Rates} and again in our concluding discussion), to make the origin of
any apparent inconsistencies fully transparent.

\optional{
For example, the inclusion of tides (cf. Hurley et al) 

- we adopt a specific max NS mass,

- HCE or not?  

: mention also physics refs?  \cite{lrr-2006-6}; see HPT; see others
}

Despite providing a distribution, our fiducial results conservatively do not account for any other  channels for forming compact objects besides
isolated binary evolution, even though several observational, semianalytic, and numerical studies have demonstrated that interacting
stellar environments can efficiently form compact binaries.
For example, in young very massive stellar clusters, a high-mass black hole subcluster can mass segregate and rapidly
form and eject merging black hole binaries  (see,e.g., \cite{PZMcM,clusters-2005,2010MNRAS.402..371B} and
references therein).   A similar process occurs in nuclear star clusters; see \cite{clus-NuclearClusters-Mergers-Miller2008}.
Alternatively, young dense clusters could undergo runaway stellar collisions, eventually forming an
``intermediate-mass'' black hole $M\in 100-10^3 M_\odot$, each of which can capture stellar-mass compact objects
\citep{gw-astro-imbh-imri-2009} 
or even form binaries and merge themselves \citep{imbhlisa-2006,gwastro-imbh-ComparableMergers-Ilya-2009}.
Theory aside,  observations suggest pulsar and X-ray binaries are highly overabundant in clusters; see, e.g.,
\cite{2000ApJ...532L..47R,2003ApJ...591L.131P,2005Sci...307..892R} and references therein.  
Further, several authors have argued observations of short GRBs' hosts, luminosities,  and redshift distribution support
a multi-component origin, as cluster dynamics lead to a different merger delay-time distribution $P(>t)$; see \cite{2006ApJ...643L..91H},
\cite{sgrb-clusters-comparison-multicomponent-Salvaterra2007},
\cite{2009AA...498..329G} and references therein.
Though present-day massive stellar clusters have little mass,  if a sufficient fraction of all star formation occurs through
clusters and a sufficient fraction occurs in massive clusters, similar highly efficient dynamical processes could produce
many merging compact binaries of all types; see, e.g., \cite{2008ApJ...676.1162S}.
To summarize, both theory and several different types of observation suggest that all types of dynamically-produced
compact binary merger exist.  These mergers could occur as  and potentially orders of magnitude more frequently than
mergers produced from field binaries;  the estimates shown in  Fig \ref{fig:ligo:rates} therefore
\emph{underestimate} the total merger rate, summed over all channels.  Nonetheless, we anticipate that future
observations might distinguish between the relative contributions of each component.  Assuming even rough separation,
our results determine how and how precisely future measurements of compact object merger rates (through
gravitational waves, short GRBs, or galactic observations) will constrain binary evolution.

\section{Conclusions and discussion}
\label{sec:Conclusions}
In this paper, we provide the first estimate of the CBC detection
rate for initial and advanced
LIGO which accounts for  (i) star formation in both elliptical and spiral galaxies as a function of redshift, (ii) binary black holes,
 (iii) a range of plausible binary fractions, and most significantly
 (iv) a large range of plausible binary evolution scenarios. 
Our results strongly indicate that  that existing gravitational wave detectors
are at the threshold of detection; that  moderate improvements
in these detectors, such as the enhanced LIGO upgrade of 2009, could plausibly lead to 
 detecting at least one binary merger in
a year of operation; 
that advanced LIGO detectors should be reasonably expected to detect inspirals over 1-2 yr of operation;
and that even the absence of a detection  by advanced interferometers would very significantly constrain the set
of model parameters  that could
be consistent with observations.

Our predicted detection rates are somewhat higher than those presented in \abbrvPSmoreconstraints{} because we include elliptical galaxies:
mergers in elliptical galaxies likely dominate our estimates of the BH-BH and BH-NS event rates.   However, our previous
estimates effectively assumed a high present-day star formation rate per unit volume: an assumed Milky Way-equivalent
galaxy density of $ 0.01/\unit{Mpc}^3$ and star formation rate per spiral galaxy of $3.5 M_\odot/\unit{yr}$ correspond to a
present-day star formation rate three times higher than observed.  For this reason, a naive product of the merger
rates per Milky Way galaxy  drawn from  \abbrvPSmoreconstraints{}'s Figure 6 leads to preferred merger rates per unit
volume that agree well our preferred values.
For example, our current results  and this naive rescaling of \abbrvPSmoreconstraints{} results both favor local merger rate densities of
$10^{-2} \unit{Mpc}^{-3}\unit{Myr}^{-1}$  for black hole-neutron
star binaries and 
$ 3\times 10^{-2} \unit{Mpc}^{-3}\unit{Myr}^{-1}$  for binary neutron
stars. 
Correcting for the star formation history normalization, however, our calculations suggest three times as many mergers
per volume for BH-NS and NS-NS than in  \abbrvPSmoreconstraints{}.  
%
Similarly, our predicted detection rates are higher than those implied by Figure 7 in \abbrvPSmoreconstraints{}, correcting for the
previously mentioned star formation rate bias (reducing the rates shown in that figure by $3$) and rescaling to a
single-interferometer rather than network range (reducing the rate by $\times 10$).

Limiting attention to elliptical- plus spiral-galaxy models that include   only those spiral galaxy history models that  reproduce 
 current observations of double pulsars, we find  that a single detector in the  initial   LIGO network  should
be sensitive to
\textbf{$2.4\times 10^{-2}$--$0.46$} $(\snrcut/\snrcutPreferred)^{-3}(D_{bns}/14\unit{Mpc})^3$
mergers per year (90\% confidence level),
where we sum over all merger types to produce
an overall detection rate, where $\snrcut=8$ is the threshold SNR of the search and
$D_{bns}$ is the radius of the effective volume inside of which a single detector in the
appropriate array could observe the inspiral of two  $1.4 M_\odot$
neutron stars.  
We estimate that the LIGO detector network   has a probability
$P_{\text{detect}}(\ge 1)\ge 0.4 + 0.5 \log (V/V_c) (T/8\unit{yr})$ of detecting a merger duing
each  year of operation, where $V=4\pi \Cv^3/3$ is the volume to which a
multidetector search for binary neutron star inspiral is sensitive,
$V_c$ is the volume inside a  $14\unit{Mpc}$ radius sphere, and $1/3<(V/V_c)(T/\unit{yr})<8$.  For example, 
 the probability
$P_{\text{detect}}$ of detecting one or more mergers can be approximated by 
(i) $P_{\text{detect}}\simeq 0.4+0.5\log
(T/0.01\unit{yr})$, assuming $\Cv=197 \unit{Mpc}$   and  it operates for $T$ years, for $T$ between $2$ days and
$0.1\unit{yr}$); or by (ii) 
$P_{\rm detect}\simeq 0.5 + 1.5 \log \Cv/32\unit{Mpc}$, for
one year of operation and for $\Cv$ between 20 and 70 Mpc. 

In this paper  we have continued to employ
approximate waveform and detection models to estimate  the
 sensitivity of ground-based gravitational wave detectors to
 moderate-mass  ($M\simeq 10 M_\odot$) BH-BH mergers (see, e.g.,
 \citet{1998PhRvD..57.4535F}  and \citet{2003PhRvD..67j4025B} for a
 review).   
Advanced detectors, however, will be sensitive at
cosmological distances to the later phases of high-mass  mergers, a complex strong-field and strongly-precessing
regime which is only beginning to be
thoroughly understood through numerical simulations.  Progress in numerical waveform modeling and search pipleline analysis
is required to better understand how often these candidates could be identified, especially for spinning systems, as
expected in nature.

\subsection{Anticipated constraints from future observations} 
Despite the existence of two competing scenarios for compact binary formation,  in many cases  the electromagnetic and
gravitational wave signatures of binaries formed dynamically and in isolation can be distinguished.  For example, black
hole binaries formed dynamically should have random spins and therefore (generically) beating gravitational inspiral
waveforms, while isolated binary evolution models favor spin-orbit alignment.  
Once binaries with isolated progenitors are identified, their  gravitational wave signals will encode both the number and properties of binaries formed through
isolated evolution, information that in turn can tightly constrain isolated binary progenitor parameters.  
To use an
extreme example, future BH-BH observations could imply a gravitational-wave detection rate $R_D$ consistent with only the highest event rates shown in
Fig. \ref{fig:ligo:rates}.   In the context of the models presented here, such an observation would imply  (i) that
elliptical galaxies host most BH-BH binaries; (ii)  that massive stars have a high binary fraction;  
(iii) that a certain phenomenological parameter  is very small ($\alpha\lambda<0.1$; \S \ref{sec:constraints}); and thus
(iv) since Milky Way observations of binary pulsars exclude low $\alpha\lambda$, that isolated binaries formed in elliptical
galaxies evolve differently to those in spirals.
However, by analogy with the insights provided from observations of isolated Galactic pulsars, typical detection rates
will be compatible with many plausible progenitor models; see  \abbrvPSmoreconstraints{}
and \cite{PSconstraints}.  Rather
than constraining single parameters tightly, the information provided by the first few observations will be encoded only in high-order correlations among many model parameters; see, e.g., Fig. 5 in \abbrvPSmoreconstraints{} and the highly
  correlated merger rate fit presented in Appendix B in \cite{PSconstraints}. 
Finally, what we can learn about binary evolution depends  on the merger detections that nature provides.  Because correlations
between parameters change significantly across the model space,
we cannot describe posterior uncertainties  once and for all.

If neutron star mergers occasionally have electromagnetic counterparts (i.e., short GRBs), then 
well-localized electromagnetic counterparts to binary
  neutron star mergers, such as (prompt or orphan) short GRB afterglows \cite{2009ARAA..47..567G,ptf-ScienceCase2009,2009arXiv0902.1527B}
  or radio
bursts,  allow optical follow-up studies of its host galaxy's  star-forming and chemical enrichment history.  Combined
with the position of the burst on the galaxy (kick) and, if within range of gravitational wave detectors, the binary
component masses, this host information can  tightly limit  isolated binary progenitor models; see, e.g., 
\abbrvPSgrbs{}.  For example, these observations could provide unique direct probes on the distribution of delays between binary birth
and merger; see, e.g., \cite{NakarReviewArticle2006},  \cite{2006ApJ...642..989P}, \cite{2007ApJ...665.1220Z}, \cite{2010ApJ...709..664R} and
references therein.  
Better still, offsets from host galaxies might provide an independent constraint on delay times (and on  supernova kick
velocities); see, e.g., \cite{ChrisShortGRBs}, \cite{2008MNRAS.385L..10T} and references therein.
Unfortunately, purely electromagnetic observations of short GRBs cannot distinguish between multiple source populations,
such as cluster-formed mergers or even non-merger-powered bursts
\citep[see,e.g.][]{grb-short-modelfit-Virgili2009,sgrb-analysis-TwoPopulationModel-Chapman-2008}.
Worse,  given plausible kick offsets, purely electromagnetic observations may not isolate the appropriate
host \citep{2009ApJ...705L.186Z}.
Fortunately, if existing space-based and ground-based resources are applied in the advanced LIGO epoch,  coincident gravitational-wave and
electromagnetic observations can resolve  these ambiguities.

Finally, as always with binary evolution, any further observations pertaining to binary stars can improve our
understanding and limit model-building freedom, such as SNIa rates;  galactic pulsar and X-ray binaries;  and even galactic WD binaries [see, e.g., the extensive discussion in \cite{2009astro2010S.221N}].

\clearpage
\begin{acknowledgements}
The authors appreciate helpful comments received from Ilya Mandel, Alberto Vecchio, Patrick Brady,
Chad Hanna, Steve Fairhurst, Tania Regimbau, Alan Weinstein, and all the members of LIGO inspiral search group.
\textbf{Grants}
R.O. was supported by National
  Science Foundation awards PHY 06 -53462 and the Center for
  Gravitational Wave Physics. The Center for Gravitational Wave
  Physics is supported by the George A. and Margaret M. Downsbrough
  Endowment and by the National Science Foundation under cooperative
  agreement PHY 01-14375. 
VK was supported by NSF grant PHY-0653321. KB acknowledges the support from the Polish Ministry
of Science and Higher Education (MSHE) grant N N203
302835 (2008-2011).
\end{acknowledgements}

\bibliographystyle{astroads}
\bibliography{%
popsyn,popsyn-sed,popsyn_gw-merger-rates,popsyn-pulsars,%
star-formation-history,star-formation-properties,%
star-evolution-theory,%
short-grb,short-grb-mergermodel,short-grb-data,short-grb-data-analysis,grb-reviewarticles,%
observations-pulsars,observations-pulsars-kicks,observations-supernovae,supernovae-theory-kicks,%
observations-galaxies-distributions,%
observations-bh-stellar,%
galaxy-formation-theory,%
structureformation-firststars,
observations-clusters,%
gw-astronomy-detection,%
gw-astronomy-mergers,gw-astronomy-mergers-approximations,%
gw-astronomy-mergers-nr,gw-astronomy-mergers-emri,%
gw-astronomy-mergers-wd,
gw-astronomy-bursts,%
gr-nr-vacuum,%
gr-nr-vacuum-kicks,%
mm-general,Astrophysics,%
astrophysics-stellar-dynamics-theory,%
astrophysics-stellar-dynamics-theory-smbhEnvironment,%
proposals-whitepapers-decadal2010,%
astronomy-facilities,%
LIGO-publications,%
technical-astronomy,
cosmology}

\appendix

\section{Reconstructing the mass and time distributions}
\label{ap:Popsyn:Smoothing}

\begin{figure*}
\includegraphics[width=\textwidth]{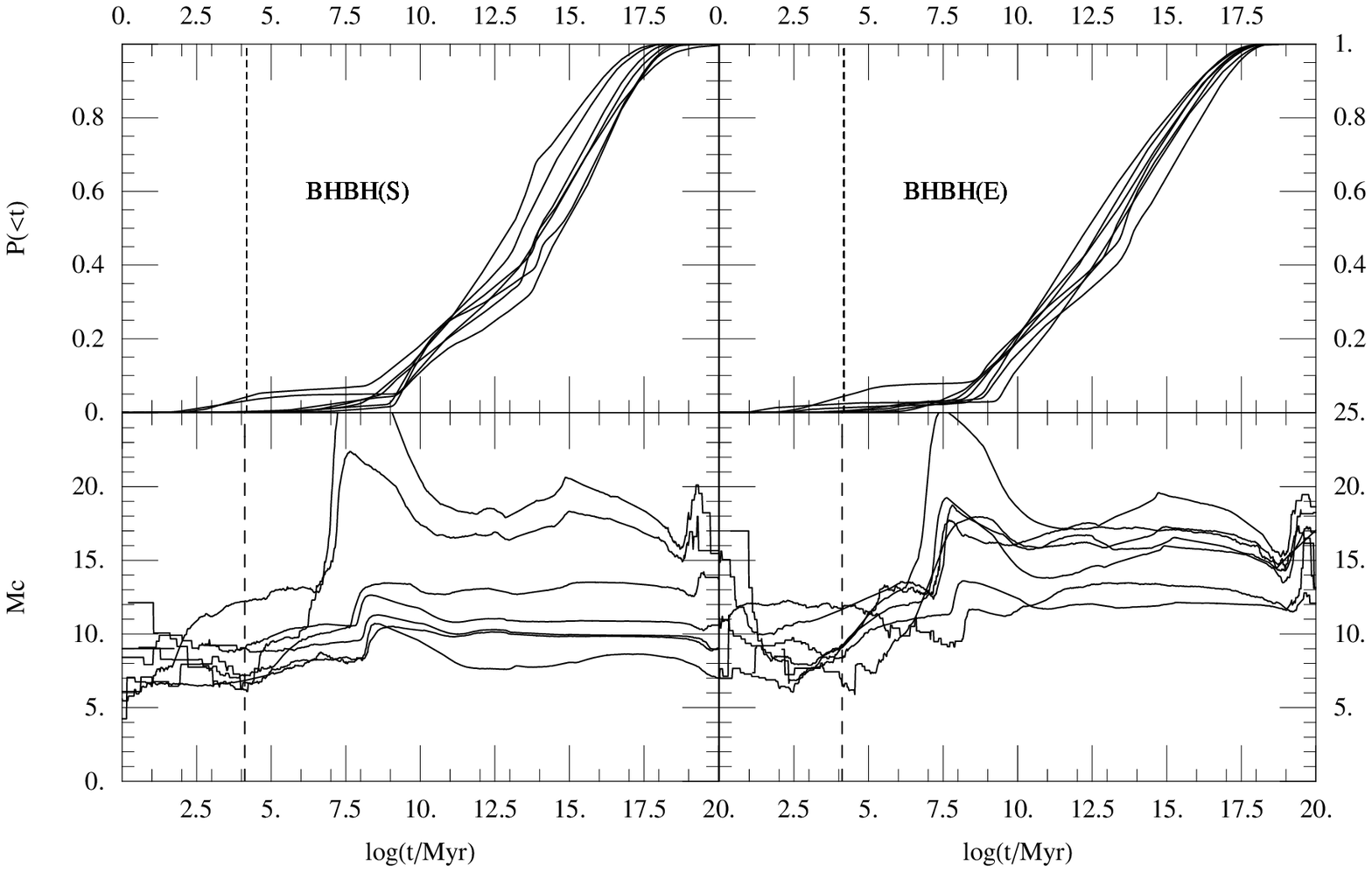}
\caption{\label{fig:popsyn:distrib:sample:pcummchirp} 
 Top panels: Cumulative probabilities
  $P_m(<t)$ that a BH-BH binary will merge in time less than $t$, for
  \textbf{ten} \emph{randomly-chosen} population synthesis models that
  satisfy the limitations discussed in the text, 
  given spiral (left) and elliptical (right) star forming conditions.
  A vertical dashed line indicates the age of the universe.
  Given the enormous sample sizes involved -- by construction, each
  sample has at least \mynmin{}  binaries -- these distributions are expected
  accurate to within \textbf{0.03} almost everywhere (with 90\% probability),
  barring the fine details near the short-time and long-time limits.
 Bottom panels: Average chirp mass $\bar{\mc}(t)$ [Eq. (\ref{eq:popsyn:mcbar})] of binaries merging between
 $0.1t$ and $10t$, versus time.   Usually, the binaries that merge
 within $100\unit{Gyr}$ after their birth have a chirp mass lower than
 $10 M_\odot$, significantly lower than the average chirp mass of more
 separated holes.
}
\end{figure*}

\noindent \emph{Selecting binaries for mass estimation}:  We prefer to smooth $dP/dt$ over a shorter timescale than the
maximum allowed by evolution of isolated BH-BH binaries.   On
physical grounds, double BH binaries whose delay times are comparable
to or smaller than the age of the universe are more likely to be
similar to merging BH-BH binaries (the focus of our study) than wide
binaries that have never interacted.  
To quantify the point at which a the transition occurs and therefore
the timescale inside of which BH-BH binaries are comparatively
similar, we calculate the ``average'' chirp mass of binaries $\tilde{\mc}(t)$
merging within a factor $10$ of  time $t$:
\begin{eqnarray}
\label{eq:popsyn:mcbar}
\bar{\mc}(t)&=&\left[ \frac{\sum_{j: 0.1 t <t_j< 10 t
     } \mc{}_j^{15/6} }{\sum_{j: t 0.1 <t_j< t
    10 } 1}
  \right]^{6/15}
\end{eqnarray}
Figure  \ref{fig:popsyn:distrib:sample:pcummchirp} shows the results
of this calculation.
Based on the delay time distribution $P_m(<t)$ and
average chirp mass of similar binaries $\bar{\mc}(t)$, a sharp
distinction exists between binaries with $t\lesssim 10^7 \unit{Myr}$
and wider binaries.  These massive wide binaries have orbits largely
unaffected by binary evolution  and have a delay
time distribution 
$dP/dt\propto 1/t$ largely determined by their initial orbital
configuration.   On the contrary, tighter and less massive binaries
have been significantly perturbed by binary evolution.  Within each
simulation, they appear to be drawn from a similar chirp mass
distribution (allowing for significant sampling-induced fluctuations in
$\bar{\mc}(t)$)

\begin{figure*}
\includegraphics[width=\textwidth]{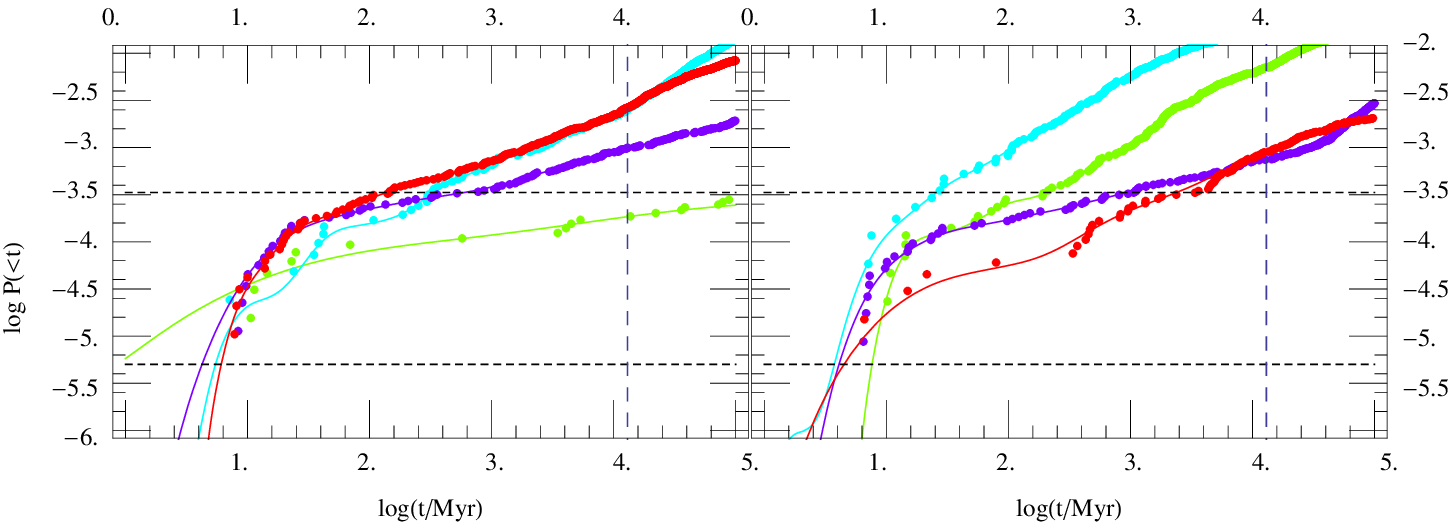}  
\caption{\label{fig:popsyn:distrib:sample:pcumTDetail}%
 Detailed view of the cumulative probabilities $P_m(<t)$ for 4 randomly selected spiral (left) and elliptical (right) models versus $\log
  t/\unit{Myr}$; compare with Figure \ref{fig:popsyn:distrib:sample:pcummchirp}.  Solid curves indicate the fit; points
  indicate the raw cumulative $P_m(<t)$.   The horizontal dashed lines indicate the smallest and largest  possibilites
  for $\log P \approx \log 1/n$, the smallest resolvable cumulative probability; see Figure \ref{fig:popsyn:selection} and the
  associated text for a discussion of its range and relation to $m$ (i.e., the largest $1/n$ occurs in only a handful of
  simulations, most of which have many merging binaries $m\gg 1$ and therefore a well-resolved $P(<t)$).  Our estimate  reproduces $P_m(<t)$ reliably at $t\simeq 100 \unit{Myr}$, the
  smallest scale on which $\rho(t)$ varies for the redshift range of interest.  Though our fits necessarily fail to  reliably recover
  $dP_m/dt$ on very short timescales,  uncertainty in integrals of $dP_m/dt$ over longer timescales average out (i.e., in  merger rate estimates ${\cal
    R}(t)$ using Eq. \ref{eq:int:comovingRate})
}

\end{figure*}


\noindent \emph{Predictions and smoothing: Physical timescales}
From the above similarity argument, we suspect the $n_{eff}$ BH-BH binaries with
delay times times $<\tcutChirpMass$ are similar to the set of binaries
with delay times $<13.5\unit{Gyr}$.  However, even
with those binaries we often have only $\sim 50$ with which to estimate
 $p(\mc)$; the chirp mass distribution must be estimated with smoothing.  We use the same approach as is used for
 $dP/dt$ in   \abbrvPSgrbs{}  to  estimate the chirp mass distribution:
\begin{eqnarray}
\label{eq:smooth:dpdmc}
\hat{p}(\mc) &\equiv&
    \frac{1}{\mc n_{eff} \ln 10}\sum_{k=1}^{n_{eff}} \frac{1}{\sqrt{2\pi(s_2)^2}}
       e^{-(\log \mc - l_{m,k})^2/2(s_2)^2} \\
s_2 &\equiv& \frac{\left[({\rm max}_k l_{m,k}) - ({\rm min}_k
    l_{m,k})\right]}{1.25\sqrt{n_{eff}}}
\end{eqnarray}
where the $l_{m,k}=\log_{10}(\mc/M_\odot)$ for $k=1\ldots n_{eff}$ are the logarithms of each binaries'
chirp mass for the $n_{eff}$ binaries with delay times $<\tcutChirpMass$.

 \noindent \emph{Absent correlations}:  We identify very few significant correlations on the timescales of interest
  between $\log \mc$ and $\log t$.  For example, the correlation coefficient between $\log \mc_k$ and $\log t_k$ for
  each simulation's 
  $n_{eff}$ close  BH-BH binaries is 
  roughly consistent with sampling error  ($\propto 1/\sqrt{n_{eff}}$). 
\optional{ In four  elliptical and
  three spiral-galaxy models, the population of $n_{eff}-m$ nearly merging BH-BH binaries (i.e., delays $13\unit{Gyr}<t<100\unit{Gyr}$)
  is marginally more massive than the population of $m$ merging binaries when $m>100$ ($\delta \log \left<\mc^{15/6}\right>^{6/15} >
  \editremark{0.09}$);  for these simulations we adopt a chirp-mass
  distribution based on the $m$ merging binaries alone.
}
For BH-NS and particularly NS-NS,
  nominally significant correlation coefficients occur.  For example, some NS-NS  $(\mc,t)$ distributions concentrate a
  fraction of mergers on and exclude a
  region below a critical $t(\mc)$  curve,  associated with mass-transfer-limited ultracompact  mergers on times
  $t\le 100\unit{Myr}$.    Other
  distributions show no apparent structure, but through  large $n_{eff}$ and therefore extremely narrow
  mass distributions yield nominally significant correlation.  Our manual inspection of all these outliers suggests no structure significant enough to
  impact merger rates (i.e., associated both with significant change in $\mc$ and long star formation timescales).
  On the contrary, as seen in Figure \ref{fig:popsyn:distrib:sample:pcummchirp}, the joint
  distribution $p(\mc,t)$ varies significantly between simulations, as the distribution in \emph{each factor} [$\mc$ and $t$] changes
  significantly.      The authors can provide
  all binary properties [$(m_1,m_2,t)$]   for each simulation on request.

\section{Sampling Limitations and Archive Selection}
\label{sec:popsyn:ArchiveSelection}

\begin{figure*}
\includegraphics[width=0.9\textwidth]{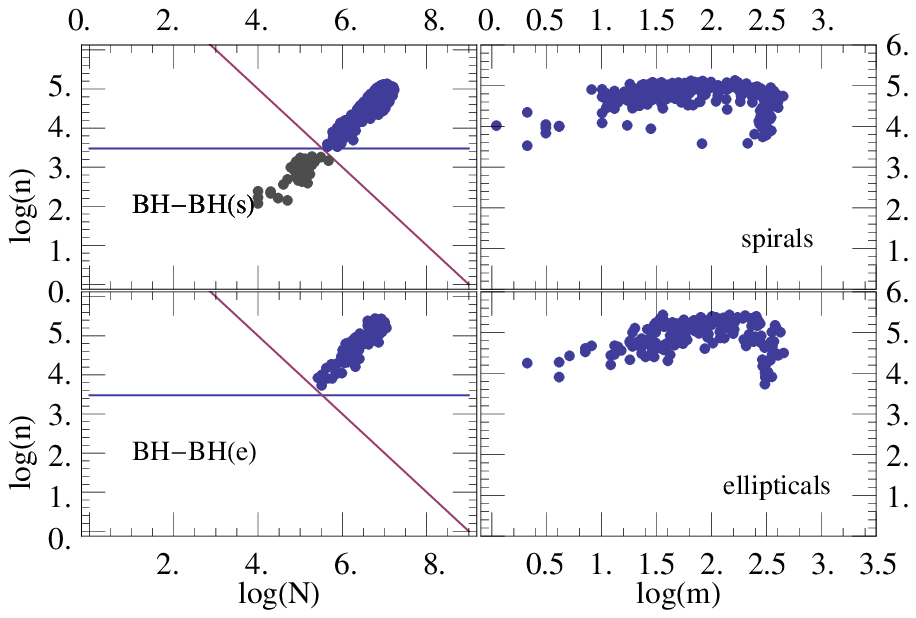}
\caption{\label{fig:popsyn:selection} 
Left plot:
For spiral (top panels) and
  elliptical (bottom) archives, a scatter plot of the number of BH-BH
  binaries $n$ seen in a simulation of $N$ progenitor binaries (each progenitor component having mass greater than $4
  M_\odot$; higher primary masses are drawn from the broken-Kroupa IMF). 
  The two solid lines show the cutoffs $n>\mynmin$ and $n N>\myratio$ imposed to
  insure data quality and reduce sampling bias.  Simulations used in
  this paper, shown in blue, must lie above and to the right of these cutoffs.
Right plot: For the simulations above and to the right of the lines
shown to left, scatter plot of
  the total number of bound BH-BH binaries remaining in the simulation
  ($n$) versus the number $m$ of \emph{potentially-merging binaries}
  (i.e., with a delay between birth and merger less than
  $13.5\unit{Gyr}$).  All selected simulations ($n N> \myratio$) have at least one merger ($m>0$).
}
\end{figure*}

Both sets of simulations are highly heterogeneous: the number of
binaries simulated ($N$), the number of bound BH-BH binaries remaining
at the end of stellar evolution ($n$), and the number of those
binaries which merge within $13.5\unit{Gyr}$ ($m$) all span several
orders of magnitude, as shown in Figure \ref{fig:popsyn:selection}.
Evidently, only for those simulations with \emph{many} BH-BH binaries ($n\gg 1000$) --
particularly with several \emph{merging} binaries ($m\gg 1$) -- can we
reliably extract relevant physics, such as the present-day merger rate
and characteristic chirp mass
of BH-BH binaries due to a starburst in the early universe.
However, as discussed in \abbrvPSgrbs, the obvious subsets of simulations for which we can reliably
make predictions, such as the set of simulations with  $n>\mynmin$, could have been
 biased, over-representing those simulations which predict more mergers per simulated binary.  
To mitigate the influence of this bias, in \emph{addition} to
requiring that each simulation have many BH-BH binaries
($n>\mynmin$), we also (or, equivalently, instead) require that $n N>\myratio$.  The second cut's prefactor ($\myratio$) is chosen so that every
simulation that satisfies it has $n>\mynmin$.
  The subset
of simulations which satisfies both conditions has a relatively
unbiased distribution of
$n/N$, the fraction of simulated binaries that end up as bound BH-BH
binaries (\abbrvPSgrbs).    
These two cuts do not introduce biases in any of the parameters: the set of simulations which satisfy
these cuts has similar statistical properties as the entire archive
(we omit the many plots needed as proof). 

\noindent \emph{Handling undersampled cases}: 
Finally, with each simulation 
guaranteed to have $\mynmin$ binaries, most simulations have a
minimum of $\sim 1$ \emph{merging} BH-BH binaries; see Figure
\ref{fig:popsyn:selection} for details.    
Some simulations still do contain very few binaries that are
merging or even nearly so, largely
because certain combinations of physical circumstances rarely lead to
the formation of many tight black hole binaries. 
Given that we construct model universes from independent ellptical and spiral simulations, these simulations will lead
to an indeterminate, small detection rate in only \orderof{$1/N_E N_S \simeq 1/300^2$ universe models}.   Even
considering only elliptical or spiral galaxy star formation, they influence our predictions \orderof{ $O(1/N_{E,S})\simeq
1\%$ of the time}.  We therefore  adopt a few percent as the lowest resolvable probability of any event our simulations
can resolve.  

\noindent \emph{Potential biases in multiply-selected simulations}:
Just as the
set of simulations was reduced to find simulations with many BH-BH
binaries in \S~\ref{sec:popsyn}, here we require our simulations
\emph{additionally} have enough NS-NS binaries to enable accurate
estimates.\footnote{We adopt the same conditions on $n$ and $N$ as
  were used in \abbrvPSgrbs.}
%
Most population synthesis simulations we  performed
that produce an adequate number of  BH-BH binaries also produce more than enough
NS-NS and NS-BH 
binaries ($\sim 10^4$) to allow us to estimate the three properties
$(\lambda,dP/dt, 
p(\mc)$ needed to reconstruct merger and detection rates for these NS
binaries.
However, 
certain combinations of population synthesis parameters  strongly favor
merging BH-BH production over NS-NS production.  Most of
our simulations of these conditions did not produce enough NS-NS
binaries to allow accurate estimates.  
In other words, the set of simulations that have \emph{both} enough
NS-NS and close BH-BH binaries is slightly biased against the highest
BH-BH and lowest NS-NS merger rates;  see Table \ref{tab:DataDump}.    

\optional{  
{\bf  
As seen in Table \ref{tab:DataDump}, these 41 spiral galaxy models split into two well-defined populations.  
One population  has large common envelope efficiency ($\alpha\gtrsim 0.7$) and empirically implausible supernova kicks
(mostly $v>500 \unit{km/s}$).
The other
population has extremely low common envelope efficiency $\alpha$ (22 out of 41 ``omitted'' simulations have $\alpha < 0.08$; cf. 32 simulations
performed with $\alpha$ in this range).   As this population correlates with low $m$ (see Table \ref{tab:DataDump}),
this population will be exclueed.

Most simulations are excluded because they produce fairly few NS-NS mergers,
compared to the strong limit on   $n N$ inherited from \abbrvPSgrbs{}.

Typically small-ish simulations ($10^6-10^7$
   binaries, with \textbf{extremely well defined} $n-m$ relation $n\simeq 1.5 m^{0.9}$ but minimal $n-N$ correlation) with only a few merging binaries of either BH-NS or NS-NS, but relatively efficient production of BH-BH
   mergers.  By eliminating them from *posterior probability* calculations and joint-PDF results, slight bias 
}
}

The goal of this paper is to determine the relative likelihood of
\abbrvCBC{} merger and detection rates given existing observations.  
In particular, in  \S~\ref{sec:constraints} we   examine
the distribution of BH-BH merger rates for (spiral-galaxy) simulations for which the NS-NS
merger rates are consistent with milky Way observations.  This subset of
simulations is inevitably contained within the biased set of
simulations for which NS-NS and BH-BH merger rates can be accurately
estimatated. 
That being said, the observational constraints reviewed in \S~\ref{sec:constraints} point to a high NS-NS
merger rate and will therefore inevitably rule out the few models for
which NS-NS merger rates could not be accurately determined.



\end{document}